\documentclass[11pt]{article}

\pdfoutput=1

\usepackage{amsmath,amssymb}
\usepackage{graphicx,color}

\numberwithin{equation}{section}

\DeclareMathOperator{\arcsinh}{arcsinh}

\DeclareMathOperator{\Tr}{Tr}

\def\({\left(}
\def\){\right)}

\newcommand{\de}{\partial}
\newcommand{\be}{\begin{equation}}
\newcommand{\ba}{\begin{eqnarray}}
\newcommand{\ea}{\end{eqnarray}}
\newcommand{\ee}{\end{equation}}

\newcommand{\f}{\frac}
\newcommand{\s}{\sqrt}

\newcommand{\ti}{\tilde}
\newcommand{\ap}{\alpha}

\newcommand{\ddd}{\cdot\cdot\cdot}
\newcommand{\no}{\nonumber \\}
\newcommand{\la}{\langle}
\newcommand{\lb}{\rangle}
\newcommand{\ep}{\epsilon}

\newcommand{\ov}{\overline}
 
 \def\de{\partial}

 \def\f {\frac}
 \def\ti{\tilde}
 \def\ap{\alpha}

 \def\ddd{\cdot\cdot\cdot}
 \def\no{\nonumber \\}

 \def\la{\langle}
 \def\lb{\rangle}
 \def\ep{\epsilon}

\begin{document}

\begin{titlepage}
\thispagestyle{empty}

\begin{flushright}
IPMU11-0136\\
MIT-CTP-4289
\end{flushright}

\vspace{.4cm}
\begin{center}
\noindent{\Large \textbf{Aspects of AdS/BCFT}}\\
\vspace{1cm}
\noindent{Mitsutoshi Fujita,\footnote[1]{e-mail: mitsutoshi.fujita@ipmu.jp}}
\noindent{ Tadashi Takayanagi\footnote[2]{e-mail: tadashi.takayanagi@ipmu.jp}}
 and \noindent{ Erik Tonni\footnote[3]{e-mail: tonni@mit.edu}}

\vspace{1cm}
{$^{1,2}$\it
 Institute for the Physics and Mathematics of the Universe (IPMU) \\
 University of Tokyo, Kashiwa, Chiba 277-8582, Japan\\
}\vspace{5mm}
{${}^{3}${\it Center for Theoretical Physics}}\\
         {\it Massachusetts Institute of Technology}\\
         {\it 77 Massachusetts Avenue, Cambridge, MA 02139, USA}\\
\vskip 2em
\end{center}

\vspace{.5cm}
\begin{abstract}
We expand the results of arXiv:1105.5165, where
a holographic description of a conformal field theory defined on a manifold
with boundaries (so called BCFT) was proposed, based on AdS/CFT. We construct gravity duals of
conformal field theories on strips, balls and also time-dependent
boundaries. We show a holographic
g-theorem in any dimension. As a special example, we can define a `boundary central charge'
in three dimensional conformal field theories and our holographic g-theorem argues that it
decreases under RG flows. We also computed holographic one-point functions and confirmed
that their scaling property agrees with field theory calculations.
Finally, we give an example of string theory embedding of this holography by inserting
orientifold 8-planes in AdS$_4\times$CP$^3$.

\end{abstract}

\end{titlepage}

\newpage

\begin{scriptsize}
\tableofcontents
\end{scriptsize}

\newpage

\section{Introduction}

The holographic principle \cite{holography} has been a very powerful idea which relates the unknown physics of quantum gravity to that of a more familiar non-gravitational theory. It also benefits us in an opposite way. We can conveniently study a strongly coupled quantum field theory by replacing it with a classical gravity theory. One of the most well studied holographic setups is the AdS/CFT correspondence \cite{Maldacena,GKP}. In most of the examples in AdS/CFT studied
so far, the holography works as an equivalence between a gravity or string theory on a
Anti de-Sitter space (AdS) and
a conformal field theory (CFT) on a compact manifold. Therefore it is intriguing to
ask what will happen if there are boundaries on the manifold on which the CFT is defined.
Such a CFT is called a boundary conformal field theory (BCFT) if a part of conformal symmetry
(called boundary conformal symmetry) is preserved at the boundaries.

\begin{figure}[bbb]
   \vspace{-.4cm}
   \begin{center}
     \includegraphics[height=4cm]{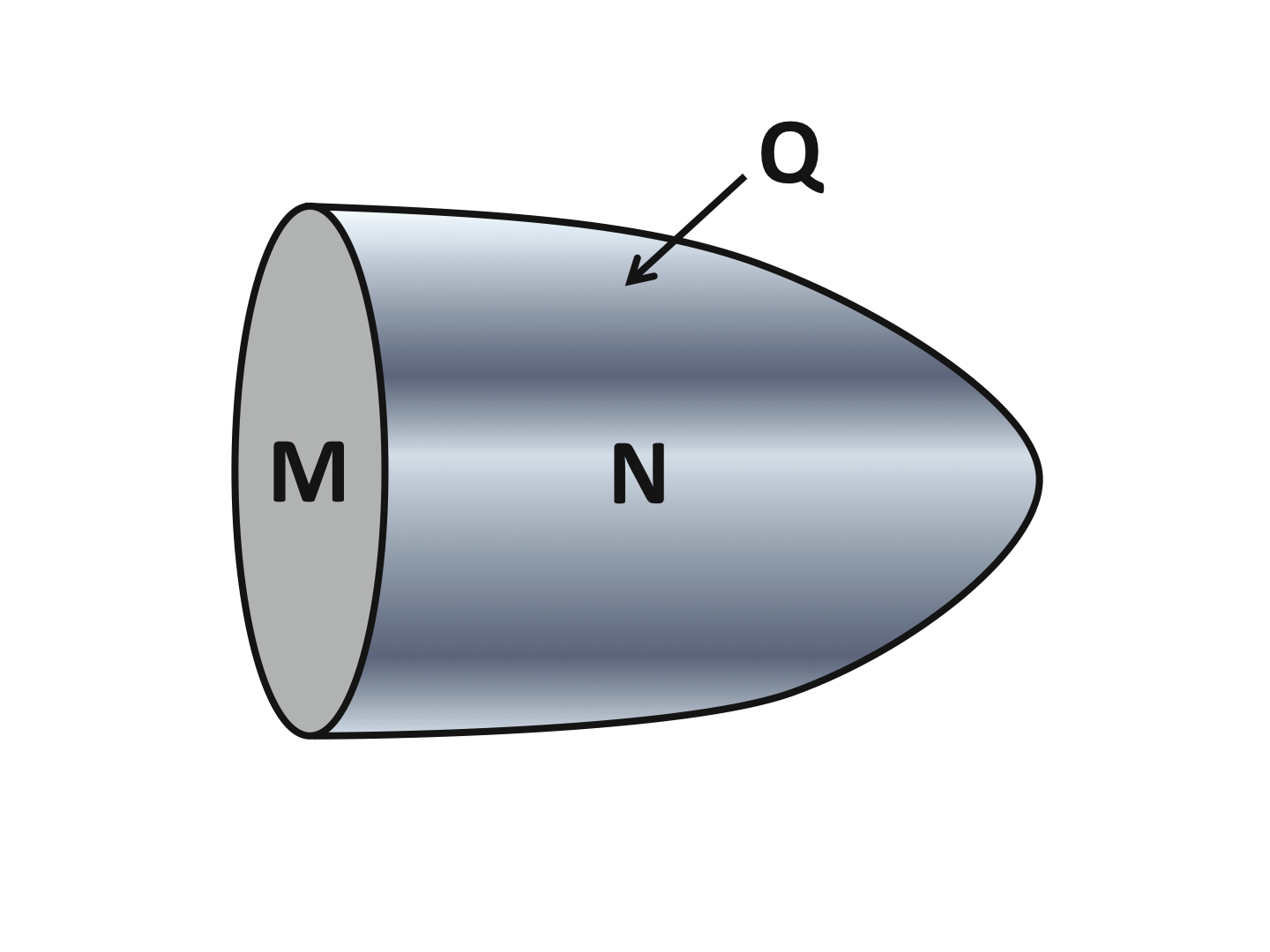}
   \end{center}
   \vspace{-.8cm}
   \caption{A schematic setup of AdS/BCFT. The CFT lives on $M$, which has the
   boundary $\de M$. Its gravity dual is denoted by $N$ and its asymptotically
   AdS is $M$. The boundary $\de M$ is extended into the bulk AdS, which
   constitutes the boundary $Q$.}\label{fig:setup}
\end{figure}

In the recent paper \cite{Ta}, an effective description of holographic dual of BCFT (AdS/BCFT)
has been generally considered and several physical quantities including the partition 
functions have been computed. We would like to note that in specific setups, such a 
holography construction of BCFT has already been mentioned in the earlier papers \cite{Kaa,Kab}.
In this construction, we introduce an extra boundary $Q$ in addition to the asymptotic AdS boundary $M$ such that the boundary of $Q$ coincides with that of $M$
(refer to Fig.\ref{fig:setup}). One of the crucial points is that we impose the Neumann boundary condition instead of the standard Dirichlet one in the gravity sector.
A new ingredient in this holography
is that we can add degrees of freedom localized at the boundary. In the two dimensional CFT,
this is measured by so called the boundary entropy or g-function \cite{AfLu}.
We would also like to mention that different constructions of holographic dual of field theories with boundaries can be found in \cite{AMR,ABBS,CDGG}.

The purpose of this paper is to further study the properties of AdS/BCFT. At the
same time, we will also expose detailed calculations on the results reported briefly in the
letter \cite{Ta}.  Some of new results in this paper are described as follows. We construct gravity duals of conformal field theories on strips, balls and also time-dependent
boundaries in any dimension. We prove a holographic
g-theorem in any dimension. As a particular example, we introduce a `boundary central charge'
in three dimensional conformal field theories. Our holographic g-theorem argues that it
decreases under RG flows. We also computed holographic one-point functions.
Finally, we give an example of string theory embedding of this holography.

This paper is organized as follows.
In section two, we review the general prescription of AdS/BCFT. In section three, we
explain the holographic calculation of boundary entropy in two dimensional conformal field theories. In section four, we study the gravity dual of a CFT on an interval and analyze
the Hawking-Page phase transition. In section five, we show the holographic g-theorem in any
dimension. In section six, we present holographic duals of CFTs defined on balls or strips
in higher dimensions. In section seven, we consider examples of AdS/BCFT with time-dependent
boundaries. In section eight, we calculate the holographic one point functions. In section nine, we present an example of string theory embedding of AdS/BCFT. In section ten, we summarize
conclusions and discuss future problems.

\section{Holographic Dual of BCFT}

We would like to formulate a holographic dual of CFT defined on a manifold $M$ with a boundary $\de M$. We argue that this is given by generalizing the AdS/CFT correspondence \cite{Maldacena}
in the following way. To have a gravity dual, we extend $d$ dimensional manifold $M$ to a $d+1$ dimensional manifold $N$ so that $\de N=M\cup Q$, where $Q$ is a manifold homologous to $M$. The conformal invariance in the bulk of $M$ requires that $N$ is a part of AdS space.

In the standard AdS/CFT, we impose the Dirichlet boundary condition at the boundary of AdS and following this we assume the Dirichlet boundary condition on $M$. On the other hand, we impose a Neumann boundary condition on $Q$ \cite{Ta}     as we will explain later. The reason for this is that this boundary should be dynamical from the viewpoint of holography and there is no
natural definite metric on $Q$ specified from the data in the CFT side. Notice that
here the boundary $Q$ is no longer asymptotically AdS. On the other hand, if we impose the Neumann boundary condition on a boundary which is parallel with the AdS boundary $M$, the setup coincides with that of the Randall-Sundrum models \cite{RaSu}. Also, we would like to note that in principle, it is also possible to adopt the Neumann boundary condition at the AdS boundary $M$ as discussed in \cite{CoMa}.

\subsection{Neumann Boundary Condition}

To make the variational problem sensible, it is conventional to add the Gibbons-Hawking boundary term \cite{GHterm} to the Einstein-Hilbert action:
\be
I=\f{1}{16\pi G_N}\int_{N}\s{-g}(R-2\Lambda)+\f{1}{8\pi G_N}\int_{Q}\s{-h}K.
\ee
The metric of $N$ and $Q$ is denoted by $g$ and $h$. $K=h^{ab}K_{ab}$ is the trace of extrinsic curvature.
The extrinsic curvature $K_{ab}$ is defined by
\be
K_{ab}=\nabla_a n_b,
\ee
where $n$ is the unit vector normal to $Q$ and here we implicitly assume a projection onto $Q$ from $N$.
For example, in the coordinate system (so called Gaussian normal coordinate)
\be
ds^2=dr^2+h_{ab}dx^adx^b,
\ee
we can explicitly show that
\be
K_{ab}=\f{1}{2}\f{\de h_{ab}}{\de r}.\label{formulak}
\ee

Now let us consider the variation of metric in the above action. After a partial integration, we find
\be
\delta I=\f{1}{16\pi G_N}\int_{Q}\s{-h}\left(K_{ab}\delta h^{ab}-Kh_{ab}\delta h^{ab}\right).
\ee
Notice that the terms which involve the derivative of $\delta h_{ab}$ cancel out thanks to the boundary term.
In this way, by imposing the Neumann boundary condition instead of the Dirichlet one, we obtain the boundary
condition
\be
K_{ab}-h_{ab}K=0. \label{boundary}
\ee

It is also possible to add some matter fields localized on $Q$ and consider a generalized action by adding
\be
I_{Q}=\int \s{-h}L_{Q}.
\ee
This modifies (\ref{boundary}) into
\be
K_{ab}-h_{ab}K=8\pi G_N T^{Q}_{ab}, \label{bein}
\ee
where we defined
\be
T^{Qab}=\f{2}{\s{-h}}\f{\delta I_Q }{\delta h_{ab}}. \label{matbc}
\ee

Before we go on, we would like to briefly remind us of the standard treatment of the
asymptotically AdS boundary $M$. We introduce the same Gibbons-Hawking boundary term
on another boundary $M$ as usual. In this case we do not need to require
(\ref{matbc}) and instead we impose the Dirichlet boundary condition $\delta h_{ab}=0$, regarding the Brown-York tensor \cite{BrYo}
\be
\tau_{ab}=\f{1}{8\pi G_N}(K_{ab}-h_{ab}K)-T^{M}_{ab}
\ee
as the holographic energy stress tensor \cite{BaKr,MyS,Sk}.

\subsection{Construction of AdS/BCFT}

As a simple class of examples, we would like to assume that the boundary matter lagrangian
$L_Q$ is simply a constant. This leads us to consider the following action (we omit the boundary terms for the AdS boundary $M$)
\be
I=\f{1}{16\pi G_N}\int_{N}\s{-g}(R-2\Lambda)+\f{1}{8\pi G_N}\int_{Q}\s{-h}(K-T).\label{act}
\ee
The constant $T$ is interpreted as the tension of the boundary $Q$. In AdS/CFT, a $d+1$ dimensional AdS space (AdS$_{d+1}$)
is dual to a $d$ dimensional CFT. The geometrical $SO(2,d)$ symmetry of AdS is equivalent to the
conformal symmetry of the CFT. When we put a $d-1$ dimensional boundary to a $d$ dimensional CFT such that the presence of the boundary breaks $SO(2,d)$ into $SO(2,d-1)$, this is called a boundary conformal
field theory (BCFT) \cite{Cbcft}. Note that this symmetry structure looks the same as that in the holography for defect or Janus CFTs \cite{Kaa,DeWolfe:2001pq,Bachas,Janus}.

The boundary condition (\ref{bein}) for the system (\ref{act}) leads to
\be
K_{ab}=(K-T)h_{ab}.\label{eqbein}
\ee
By taking its trace, we obtain
\be
K=\f{d}{d-1}T.
\ee

To realize expected symmetries, we employ the following
ansatz of the metric
\be
ds^2=d\rho^2+\cosh^2\f{\rho}{R}\cdot
ds^2_{AdS_{d}}.
\label{metads}
\ee
If we assume that $\rho$ takes all
values from $-\infty$ to $\infty$, then (\ref{metads}) is equivalent
to the AdS$_{d+1}$. To see this, let us assume the Poincare metric
of AdS$_d$ by setting \be
ds_{AdS_d}^2=R^2\f{-dt^2+dy^2+d\vec{w}^2}{y^2},\label{metadss} \ee
where $\vec{w}\in \mathbb{R}^{d-2}$. Remember that the cosmological constant
$\Lambda$ is related to the AdS radius $R$ by \be
\Lambda=-\f{d(d-1)}{2R^2}. \ee If we define new coordinates $z$ and
$x$ by \be z=\f{y}{\cosh\f{\rho}{R}},\ \ \ \ \ \ \
x=y\tanh\f{\rho}{R}, \label{corc} \ee then we obtain the familiar form of the
Poincare metric of AdS$_{d+1}$ \be
ds^2=R^2\f{dz^2-dt^2+dx^2+d\vec{w}^2}{z^2}.\label{poin} \ee

\begin{figure}[ttt]
   \begin{center}
     \includegraphics[height=4cm]{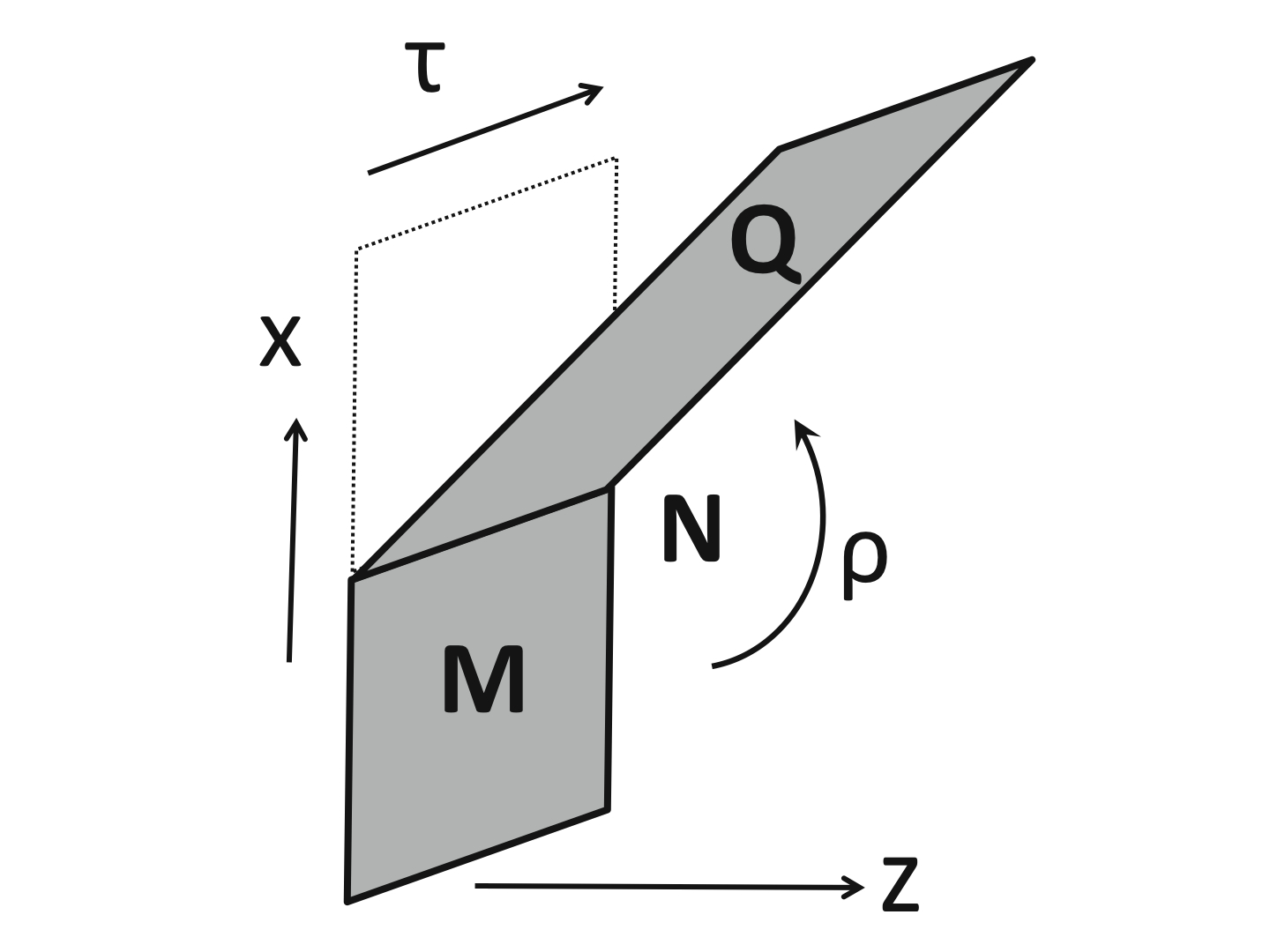}
   \end{center}
   \vspace{-.4cm}
   \caption{The holographic dual of a half line. The spacetime  dual to BCFT is restricted to the region $-\infty <\rho <\rho_*$ and is surrounded by the shaded region.}\label{figline}
\end{figure}

Now to realize a gravity dual of BCFT, we will put a boundary $Q$ at $\rho=\rho_*$ and this means that
we restrict the spacetime to the region $-\infty<\rho<\rho_*$ as depicted in Fig.\ref{figline}.
The extrinsic curvature on $Q$ can be found from (\ref{formulak})
\be
K_{ab}=\f{1}{R}\tanh\left(\f{\rho}{R}\right)h_{ab}.
\ee
By using (\ref{eqbein}),    $\rho_*$ is determined by the tension $T$ as follows
\be
T=\f{d-1}{R}\tanh\f{\rho_*}{R}.\label{tension}
\ee
This leads to the constraint $-(d-1)\leq TR\leq d-1$.

\section{AdS$_3/$CFT$_2$ and Boundary Entropy}

Here let us concentrate on the case of $d=2$, which describes the two dimensional BCFT.
The reason that this setup is special is that it has been well studied in the subject of
two dimensional CFT \cite{Cardy}. Moreover, the BCFT has an interesting quantity called the boundary entropy (or $g$-function) introduced in \cite{AfLu}.

The boundary state of a BCFT with a boundary condition $\ap$
is denoted by $|B_\alpha\lb$ below. The function called $g$ is defined by the disk
amplitude \cite{AfLu}
\be
g_\ap=\la 0|B_\alpha\lb,
\ee
where $|0\lb$ is the vacuum state. The boundary entropy $S^{(\ap)}_{bdy}$ is simply defined by
\be
S^{(\ap)}_{bdy}=\log g_{\ap}. \label{gfun}
\ee
Below we will present two different calculations of the boundary entropy, which turns
out to coincide as expected. Later, in section \ref{ghigh} we will provide a third independent
holographic calculation of the boundary entropy.

\subsection{Boundary Entropy from Disk Partition Function}\label{sec:be}

\begin{figure}[ttt]
   \begin{center}
     \includegraphics[height=4cm]{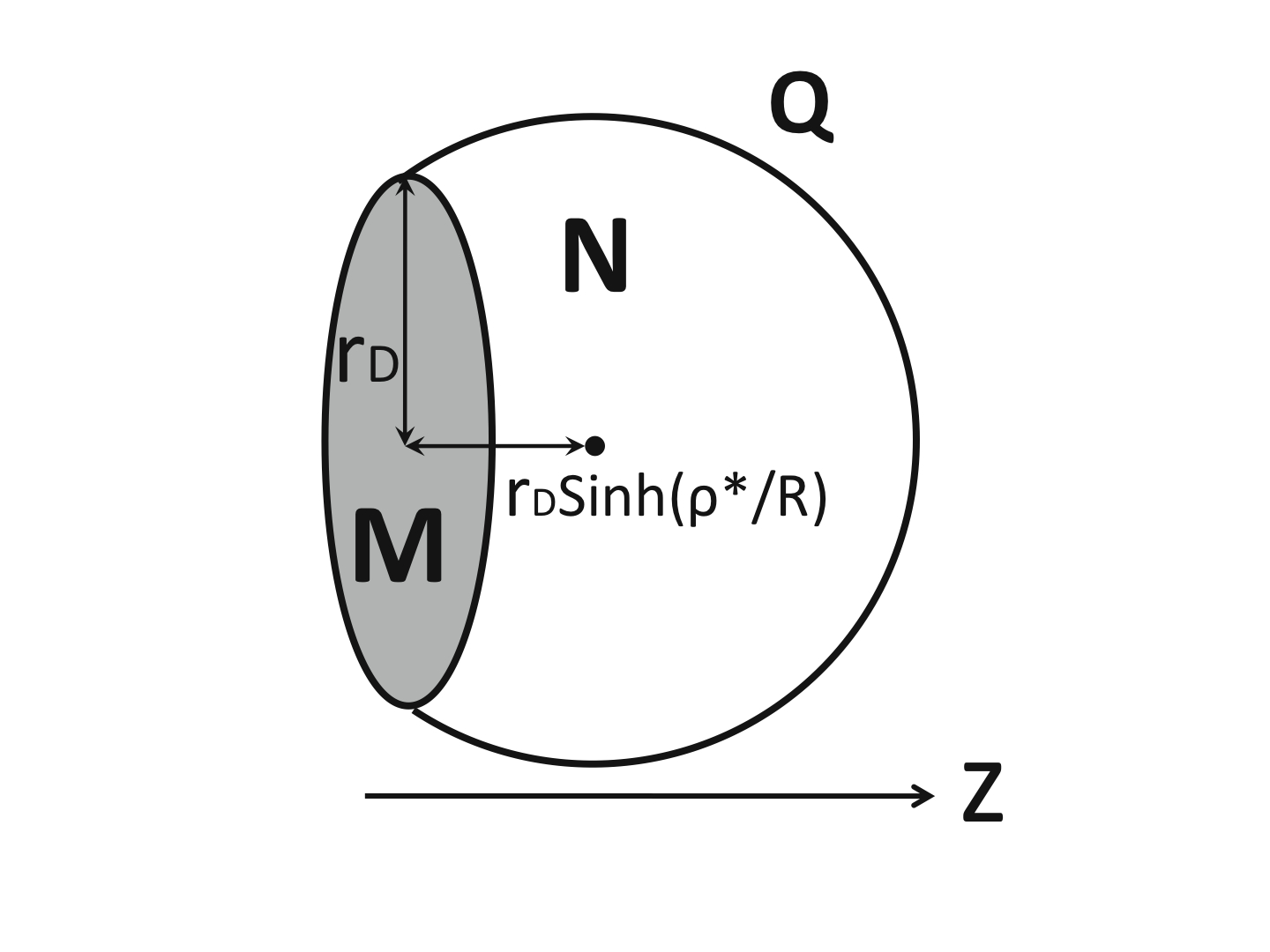}
   \end{center}
   \vspace{-.8cm}
   \caption{The holographic dual of a disk.}\label{figsphere}
\end{figure}

Consider a holographic dual of a CFT on the round disk defined by $t^2_E+x^2\leq r^2_D$ in the
Euclidean AdS$_3$ spacetime
\be
ds^2=R^2\f{dz^2+d\tau^2+dx^2}{z^2},\label{poth}
\ee
where $\tau$ is the Euclidean time. In the Euclidean formulation, the action (\ref{act})
is now replaced by
\be
I_E=-\f{1}{16\pi G_N}\int_{N}\s{g}(R-2\Lambda)-\f{1}{8\pi G_N}\int_{Q}\s{h}(K-T).\label{acte}
\ee
Note that $\rho_*$ is related to the tension $T$ of the boundary via (\ref{tension}).
When the BCFT is defined on the half space $x<0$, its gravity dual has been found in previous section.
Therefore we can find the gravity dual of the BCFT on the round disk by applying the
conformal map
\ba
&& x'_\mu=\f{x_\mu+cx^2}{1+2(c\cdot x)+c^2\cdot x^2},\no
&& z'=\f{z}{1+2(c\cdot x)+c^2\cdot x^2},  \label{cmap}
\ea
where $c^\mu=(c^\tau,c^x)$ are arbitrary constants \cite{BM}.
Finally, we obtain the following domain in AdS$_3$
\be
\tau^2+x^2+\left(z-\sinh(\rho_*/R)r_D\right)^2-r_D^2\cosh^2(\rho_*/R)\leq 0. \label{diskh}
\ee
In this way we found that the holographic dual of BCFT on a round disk is given by a part of
the two dimensional round sphere as described in Fig.\ref{figsphere}. A larger value of tension corresponds to the larger radius.

Now we would like to calculate the disk partition function in order to obtain the boundary entropy. For this we just need to evaluate (\ref{acte}) in the domain (\ref{diskh}). In the end we find
\ba
I_E&=&\f{R}{4G_N}\left(\f{r^2_D}{2\ep^2}+\f{r_D\sinh(\rho_*/R)}{\ep}+\log(\ep/r_D)-\f{1}{2}
-\f{\rho_*}{R}\right), \label{diskpar}
\ea
where we introduced the cutoff such that $z>\ep$. By adding the counter term on the AdS boundary, we can subtract the divergent terms in (\ref{diskpar}). The difference of the partition function
between $\rho=0$ and $\rho=\rho_*$ is given by
\be
I_E(\rho_*)-I_E(0)=-\f{\rho_*}{4G_N}.
\ee
Since the partition function is given by $Z=e^{-S_E}$, we obtain the boundary entropy
\be
S_{bdy}=-I_E=\f{\rho_*}{4G_N}=\f{c}{6}\mbox{arctanh}(RT), \label{bed}
\ee
where implicitly we assumed $S_{bdy}=0$ at $T=0$ because the boundary contributions
vanish in this case.

By employing the integrals computed to get (\ref{diskpar}), we can evaluate the action in the case of a domain $M$ in the boundary given by an annulus. We consider as $Q=Q_1 \cup Q_2$ the disconnected configuration shown in Fig. \ref{figannulus} and we discuss a connected configuration in the appendix \ref{appannulus}.\\
Since the orientation of the two surfaces $Q_1$ and $Q_2$ of the
disconnected boundaries are opposite, the definition of tension is opposite.
This means that $\rho_{*,1}>0$ at $Q_1$ corresponds to $T>0$ (or equally large
boundary entropy), while $\rho_{*,2}>0$ at $Q_2$ corresponds to $T<0$ (or smaller boundary entropy).
Thus, the result is given by the difference of two contributions like (\ref{diskpar})
\be
I_E \,=\,
\frac{R}{4 G_N}
 \left[\,\frac{r^2_{D,2}-r^2_{D,1}}{2\epsilon^2}
 +\frac{r_{D,2}\,\sinh(\rho_{\ast,2}/R)-r_{D,1}\,\sinh(\rho_{\ast,1}/R)}{\epsilon}+\log\bigg(\frac{r_{D,1}}{r_{D,2}}\bigg)
-\frac{\rho_{\ast,2}-\rho_{\ast,1}}{R}\,\right]
\ee
where the logarithmic divergence is canceled in the difference.

\subsection{Boundary Entropy from Holographic Entanglement Entropy}

Another way to extract the boundary entropy is to calculate the entanglement entropy.
The entanglement entropy $S_A$ with respect to the subsystem $A$ is
defined by the von Neumann entropy $S_A=-\Tr \rho_A\log\rho_A$
for the reduced density matrix $\rho_A$. The reduced density matrix $\rho_A$
is defined by tracing out the subsystem $B$, which is the complement
of $A$. In quantum field theories, we specify the subsystem $A$
by dividing a time slice into two regions. For a two dimensional CFT with a boundary (i.e. BCFT)
we can generally obtain the following result \cite{CaCa}
\be
S_A=\f{c}{6}\log\f{l}{\ep}+\log g_\ap,\label{entg}
\ee
where $c$ is the central charge and $\ep$ is the UV cut off (or lattice spacing);
the subsystem $A$ is chosen to be an interval with length $l$ such that it includes the boundary.

In AdS/CFT, the holographic entanglement entropy \cite{RT} is given in terms of the area of the codimension two minimal surface (called $\gamma_A$) which ends at $\de A$
\be
S_A=\f{\mbox{Area}(\gamma_A)}{4G_N}. \label{arealaww}
\ee
This calculation of boundary entropy from the holographic entanglement entropy has been first applied to Janus CFTs in \cite{BE}. Moreover, for supersymmetric Janus CFTs, the excellent agreement between the holographic result and the CFT result has been confirmed in \cite{BES}.

Consider the gravity dual of two dimensional BCFT on a half space $x<0$
in the coordinate (\ref{poth}). By taking the time slice $\tau=0$, we define the subsystem $A$
by the interval $-l\leq x\leq 0$.
In this case, the minimal surface (or geodesic line) $\gamma_A$
is given by  $x^2+z^2=l^2$. If we go back to the coordinate system (\ref{metads}) and
(\ref{metadss}), then $\gamma_A$ is simply given by $\tau=0, y=l$ and $-\infty<\rho\leq\rho_*$. Then
\be
S_A=\f{1}{4G_N}\int^{\rho_*}_{-\rho_{\infty}}d\rho=\f{\rho_*+\rho_{\infty}}{4G_N}.\label{geod}
\ee
Here $\rho_{\infty}$ is related to the UV cut off using (\ref{corc}) via
$\rho_{\infty}=R\log \f{2l}{\epsilon}$. By subtracting the bulk contribution which is divergent as in (\ref{entg}), we finally find
\be
S_{bdy}=S_A(\rho_*)-S_A(0)=\f{\rho_*}{4G_N}.
\ee
 This indeed agrees with (\ref{bed}).

 Actually, in the formula (\ref{arealaww}), we need to choose the end point of the
 geodesic $\gamma_A$ on $Q$ such that the total length takes the minimum value.
 Indeed, we can check explicitly that this minimum is realized when $\gamma_A$ is
 a circle with radius $l$ as is assumed in (\ref{geod}).

\begin{figure}[ttt]
   \vspace{-1.8cm}
   \begin{center}
     \includegraphics[height=9cm]{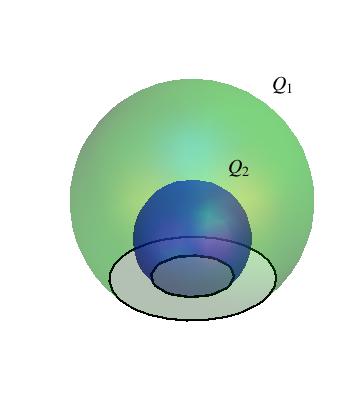}
   \end{center}
   \vspace{-1.8cm}
   \caption{The holographic dual of an annulus made by two disconnected surfaces in the bulk.}
   \label{figannulus}
\end{figure}

\section{Holographic Dual of Intervals and Hawking-Page Transition}

So far we studied the holographic dual of BCFT in the presence of a single boundary.
As a next step, we would like to analyze the holographic dual of two dimensional CFT on an interval in the setup of
AdS$_3/$CFT$_2$ as one of the simplest examples with multiple boundaries. We assume such a
system at finite temperature and there are two candidates for the bulk geometry, one of them is the thermal AdS$_3$
and the other is the BTZ black hole (AdS$_3$ black hole) \cite{BTZ}.
In the absence of the boundaries, the former is favored
at low temperature, while the latter at high temperature. It is natural to expect a similar phase structure and indeed we will confirm this below.

\subsection{Low Temperature Phase}

At low temperature, the bulk geometry is expected to be given by the thermal AdS$_3$ defined by
the metric
\be
ds^2=R^2\left(\f{d\tau^2}{z^2}+\f{dz^2}{h(z)z^2}+\f{h(z)}{z^2}dx^2\right),
\label{tads}
\ee
where $h(z)=1-(z/z_0)^2$. The periodicity of the Euclidean time $\tau$, denoted by
the inverse temperature $1/T_{BCFT}(\equiv 2\pi z_H)$, can be chosen arbitrary, while that of the space direction $x$ is determined to be $2\pi z_0$ by requiring the smoothness.

We can describe the boundary $Q$ by the curve $x=x(z)$.
The space-like unit vector $n^\mu$ normal to the surface $Q$ is given by
\be
(n^\tau,n^z,n^x)=(0,-x'(z)h(z)^2,1)\cdot \f{z}{R\s{h(z)(1+h(z)^2x'(z)^2)}}.
\ee
The extrinsic curvature $K_{ab}$
can be computed by following the procedure explained in the
appendix \ref{apext}. In the end, the boundary condition (\ref{eqbein})
leads to the following constraint on the profile of $Q$
\be
\f{dx}{dz}=\f{RT}{h(z)\s{h(z)-R^2T^2}},   \label{tadsx}
\ee
which is solved (fixing the constant shift by setting $x(0)=0$)
\be
x(z)=z_0\cdot \arctan\left(\f{RTz}{z_0\s{h(z)-R^2T^2}}\right).  \label{thsx}
\ee

\begin{figure}[ttt]
   \begin{center}
     \includegraphics[height=5cm]{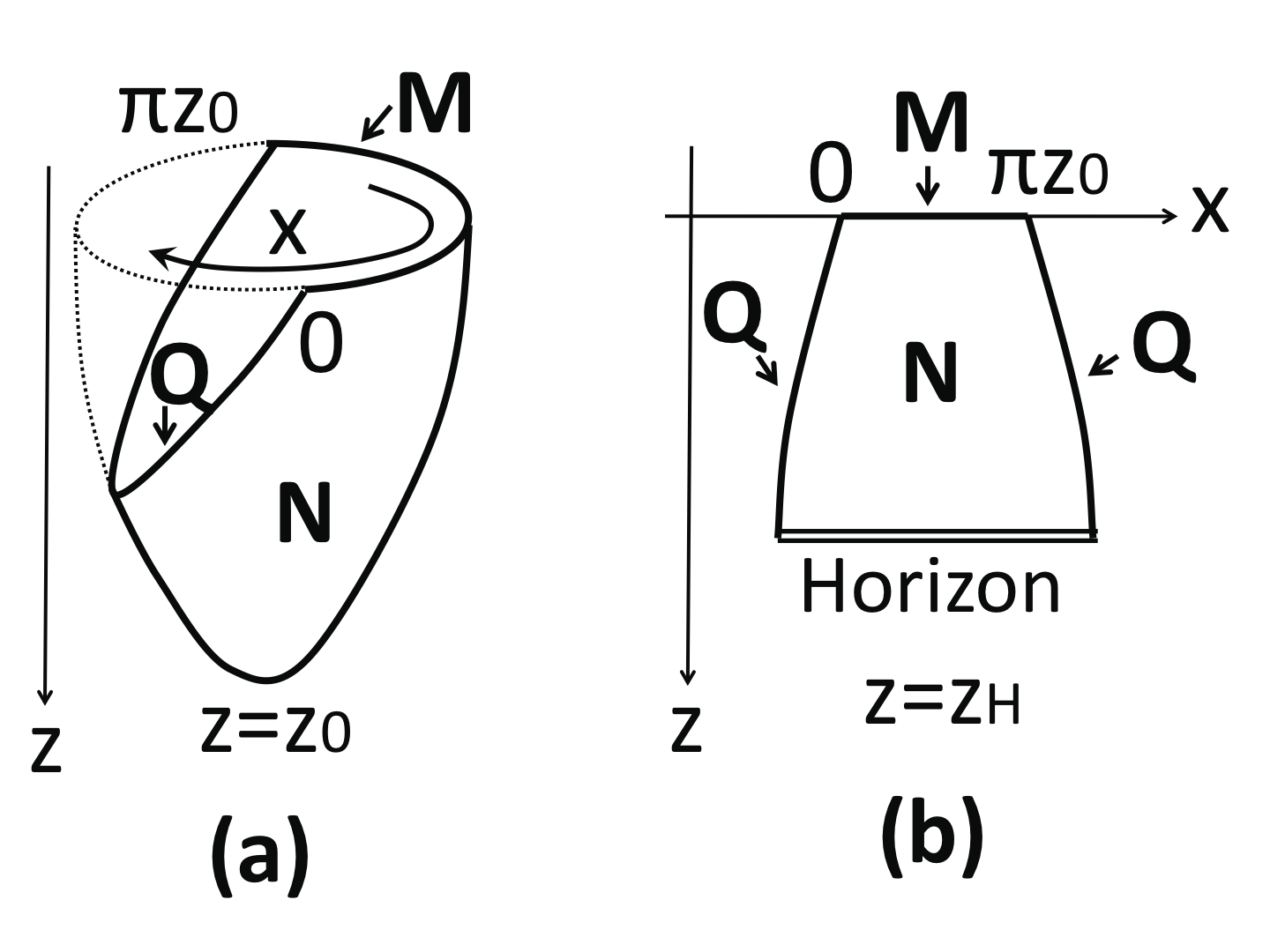}
   \end{center}
   \vspace{-1cm}
   \caption{The holographic dual of an interval at low temperature
   (a) and high temperature (b).}\label{fig:transition}
\end{figure}

Notice that $x'(z)$ gets divergent at $z_*=z_0\s{1-R^2T^2}$ and thus this should be the turning point. Thus the boundary $Q$ extends from $x=0$ to $x=\pi z_0$. Assuming $T>0$, the bulk spacetime $N$ is defined by the sum of
$(-\pi z_0\leq x\leq 0,\  0<z\leq z_0) $ and $(0<x\leq \pi z_0,\  z(x)<z<z_0)$, where
$z(x)$ is the inverse function of (\ref{thsx}) and its extension to $\f{\pi}{2}z_0<x<\pi z_0$.
This is described in Fig.\ref{fig:transition}(a).

Now the Euclidean action (\ref{acte}) reads
\ba
I_E=\f{Rz_H}{G_N}\int^{z_*}_\ep\f{dz}{z^3}\left(x(z)+\f{\pi z_0}{2}\right)+\int^{z_0}_{z_*}
\f{dz}{z^3}(\pi z_0)-\f{z_H TR^2}{2G_N}\int^{z_*}_\ep \f{dz}{z^2\s{h(z)-R^2T^2}}, \label{tadsac}
\ea
where $\ep$ is the UV cut off as before.
To evaluate (\ref{tadsac}) by eliminating the divergence, we need to be careful in that we have to regard $2\pi \ti{z}_0$ as the physical radius, defined by
\be
\ti{z}_0=\s{f(\ep)}\,z_0, \label{adj}
\ee
by matching the asymptotic geometry at $z=\ep$. Also the contribution Gibbons-Hawking term at the AdS boundary $M$ is vanishing as usual, by using the boundary integral of $K-K^{(0)}$ instead of that of $K$, where $K^{(0)}$ is the trace of extrinsic curvature for the pure AdS$_3$ (\ref{poth}). In the end, we obtain the result
\be
I_E=-\f{\pi Rz_H}{8G_Nz_0}=-\f{\pi}{24}\cdot\f{c}{\Delta x\cdot T_{BCFT}}, \label{tadsi}
\ee
where we employed the well-known relation \cite{BrHe} between the AdS$_3$ radius $R$ and
the central charge $c$ of $CFT_2$, given by
\be
c=\f{3R}{2G_N}.
\ee
Note that the final result (\ref{tadsi}) does not depend on the tension $T$ and we can confirm that (\ref{tadsi}) is also correct when $T<0$.

\subsection{High Temperature Phase}\label{ghigh}

We expect that in the higher temperature phase the bulk is described by a part of the BTZ black hole
\be
ds^2=R^2\left(\f{f(z)}{z^2}d\tau^2+\f{dz^2}{f(z)z^2}+\f{dx^2}{z^2}\right),
\ee
where $f(z)=1-(z/z_H)^2$. The Euclidean time $\tau$ is compactified on a circle such that
$\tau\sim \tau + 2\pi z_H$ and thus the temperature in the dual BCFT is $T_{BCFT}=\f{1}{2\pi z_H}$.
The length of the interval is again denoted by
$\Delta x=\pi z_0$.

We specify the boundary $Q$ by the profile $x=x(z)$.

The space-like unit vector $n^\mu$ normal to the surface $Q$ is given by
\be
(n^\tau,n^z,n^x)=(0,-x'(z)f(z),1)\cdot \f{z}{R\s{h(z)(1+f(z)x'(z)^2)}}.
\ee
The extrinsic curvature $K_{ab}$ can be again
computed by following the procedure explained in the
appendix \ref{apext}. In the end, we find that the boundary condition (\ref{eqbein}) requires
\be
\f{dx}{dz}=\f{RT}{\s{1-R^2T^2f(z)}}.   \label{btzx}
\ee
This is solved as
\be
x(z)=z_H\cdot \arcsinh\left(\f{RTz}{z_H\s{1-R^2T^2}}\right).  \label{solx}
\ee
To realize the holographic dual of the interval we need two boundaries as described in Fig.\ref{fig:transition}(b).
In the appendix \ref{aprot} we study the rotating BTZ black hole generalizing (\ref{solx}) for that case (see (\ref{Fil06})).

Now we are able to evaluate the Euclidean action (\ref{acte})
in the form
\ba
I_E=2I_{bdy}+I_{bulk},
\ea
where $2I_{bdy}$ is the contribution from the presence of two boundaries which is vanishing
if we set $T=0$. $I_{bulk}$ is the bulk contribution which does not depend on $T$.
To calculate them we subtract the divergence in a standard holographic renormalization. For the bulk part, we obtain
\be
I_{bulk}=-\f{\pi c}{6}\Delta x \cdot T_{BCFT}, \label{bulkbtz}
\ee
where we have to be careful about the issue like (\ref{adj}). This result (\ref{bulkbtz}) clearly agrees with what we expect from the
standard CFT results. On the other hand, each of two boundary contributions
is found to be
\ba
&& \f{Rz_H}{2G_N}\int^{z_H}_\ep dz \f{x(z)}{z^3}-\f{T R^2z_H}{4G_N}
\int^z_\ep \f{dz}{z^2}\f{1}{\s{1-R^2T^2f(z)}}\no
&& =-\f{Rz_H}{4G_N}\left[\f{x(z)}{z^2}\right]^{z_H}_\ep=-\f{\rho_*}{4G_N}+(\mbox{divergent terms}),
\ea
where we relate the tension $T$ to $\rho_*$ via the previous relation (\ref{tension}) and the index $\alpha$ describes each of the two boundary. Therefore we obtain the contribution to the Euclidean action from each of the boundaries
\be
I_{bdy}=-\f{\rho_*}{4G_N}=-\f{c}{6}\mbox{arctanh}(RT). \label{bdybtz}
\ee
Thus the total action is found to be
\be
I_E=-\f{\pi c}{6}\Delta x \cdot T_{BCFT}-\f{c}{3}\mbox{arctanh}(RT). \label{btzi}
\ee
The total entropy of this thermal system is found from (\ref{bulkbtz}) and (\ref{bdybtz})
\be
S_{thermal}=\f{\pi}{3}c\Delta x\cdot T_{BCFT}+\f{c}{3}\mbox{arctanh}(RT)
\ee

We would also like to point out that this calculation offers
one more different calculation of boundary entropy $S_{bdy}$ in AdS/CFT.
Consider a BCFT at a finite temperature
$T_{BCFT}$, in other words, a CFT defined on a cylinder,
the two boundary conditions imposed
on its two boundaries are denoted by $\ap$ and $\beta$. They are described by the boundary states
$|B_{\ap}\lb$ and $|B_{\beta}\lb$. The partition function $Z_{\ap\beta}$
on a cylinder, whose length is denoted by $\Delta x$, gets factorized in the high temperature limit
$T_{BCFT}\Delta x>>1$
\be
\la B_{\ap}|e^{-H\Delta x}|B_{\beta}\lb\simeq g_\ap g_\beta e^{-E_0 \Delta x}, \label{bstate}
\ee
where $H$ is the Hamiltonian (in closed string channel)
and $E_0$ is the ground state energy. The final factor $e^{-E_0\Delta x}$ is interpreted as the
thermal energy for the CFT as is clear in opens string channel. Therefore the contribution from
the presence of boundary is the product of g-function $g_\ap g_\beta$ \cite{AfLu}.
In our holographic calculation,
this means $g=e^{S_{bdy}}=e^{-I_{bdy}}$ and this is indeed true by comparing (\ref{bdybtz}) and (\ref{bed}).

\subsection{Phase Transition}

\begin{figure}[tb]
\vspace{-1.2cm}
   \begin{center}
     \includegraphics[height=6cm]{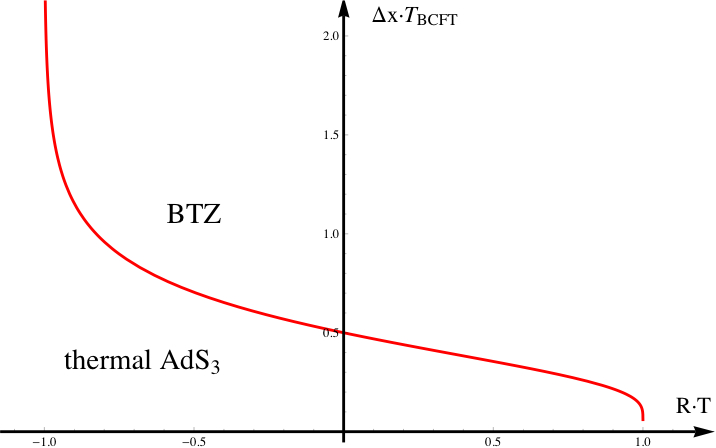}
   \end{center}
   \vspace{-.4cm}
   \caption{The plot of the critical temperature of the phase transition between
   black hole phase (the region above the curve) and thermal AdS phase (the region below the curve).}
   \label{fig:temp}
\end{figure}

Let us examine when either of the two phases is favored. To see this we compare (\ref{tadsi}) and
(\ref{btzi}) and pick up the smaller one. In this way we find that the black hole phase is
realized if and only if
\be
\Delta x\cdot T_{BCFT}>-\f{1}{\pi}\mbox{arctanh}(RT)+\s{\f{1}{4}+\f{1}{\pi^2}\mbox{arctanh}^2(RT)},
\ee
as plotted in Fig.\ref{fig:temp}. At lower temperature, the thermal AdS phase is favored.
At vanishing tension $T=0$, the phase boundary $z_0=z_H$ coincides with that of the Hawking-Page transition \cite{HaPa}.
As the tension gets larger, the critical temperature gets lower. This is consistent with the fact that the entropy
$S_{bdy}$ carried by the boundary increases as the tension does. Note also that the critical temperature gets vanishing at the maximum tension $TR=1$ and that it gets divergent at the smallest tension $TR=-1$. This phase transition is first order and can be regarded as a
generalization of the Hawking-Page phase transition dual to the
confinement/deconfinement transition in gauge theories \cite{Wi}.

\section{Holographic g-theorem}

In two dimension, the central charge $c$  is a very useful quantity which characterizes the degrees of freedom of a given CFT. Moreover, there is a well-known fact, so called c-theorem, that the central charge decreases under the RG flow. We can construct a c-function which interpolates the two central charges in two CFTs which are connected by a RG flow such that
it is a monotonically decreasing function \cite{cth}. Holographic proofs of c-theorems have been
obtained e.g. in \cite{hcth}.

In the case of BCFT, an analogous quantity is actually known and is called g-function or boundary entropy. At fixed points of boundary RG flow, they are reduced to that of BCFT as already mentioned in (\ref{gfun}). The monotonicity similar to the c function has been
conjectured in \cite{AfLu} and been shown in \cite{FrKo}. This is called the g-theorem.
Here we would like to study the holographic proof of this g-theorem. Refer to \cite{Ya} for an earlier work where the holographic c-theorem has been studied in a probe approximation.

\subsection{Holographic g-theorem in 2D CFT}

Consider the AdS$_3/$BCFT$_2$ setup and we just consider a pure gravity in the bulk as
we would like to keep the bulk conformal invariance. Since all solutions to the Einstein equation with a negative cosmological constant are locally $AdS_3$, we can assume that the gravity dual is given just by cutting a $AdS_3$ along a boundary $Q$. We describe the boundary $Q$ by the curve $x=x(z)$ in the metric (\ref{poth})
as before. We assume generic matter fields on $Q$ and this leads to the energy stress
tensor $T^Q_{ab}$ term in the boundary condition as explained in (\ref{bein}). It is easy to
check the energy conservation $\nabla^a T^{Q}_{ab}=0$ because $\nabla^a(K_{ab}-Kh_{ab})=R_{n b}\propto g_{nb}=0
$ in Einstein manifolds, where $n$ is the Gaussian normal coordinate which is normal to $Q$.

Now we would like to require that the matter fields on the boundary is physically sensible. In particular, we impose the null energy condition (or equally weaker energy condition) as is done
usually for the holographic proof of c-theorem. It is given by the following inequality for any null vector $N^a$
\be
T_{ab}^QN^aN^b\geq 0. \label{nulle}
\ee
In our case, we can choose
\be
N^t=\pm 1,\ \ \  \ N^z=\f{1}{\s{1+(x'(z))^2}},\ \ \  \  N^x=\f{x'(z)}{\s{1+(x'(z))^2}}.
\ee
Then the condition (\ref{nulle}) leads to
\be
(K_{ab}-Kh_{ab})N^aN^b=-\f{R\cdot x''(z)}{z\left(1+(x'(z))^2\right)^{3/2}}\geq 0.
\ee
Thus we obtained the condition
\be
x''(z)\leq 0.\label{cond}
\ee

Since at a fixed point the boundary entropy is given by $S_{bdy}=\f{\rho_*}{4G_N}$ and we have the relation $\f{x}{z}=\sinh(\rho_*/R)$ on the boundary $Q$, we would like to propose the following
$g$-function
\be
\log g(z)=\f{R}{4G_N}\cdot \arcsinh\left(\f{x(z)}{z}\right). \label{gfunca}
\ee
By taking derivative, we get
\be
\f{\de \log g(z)}{\de z}=\f{x'(z)z-x(z)}{\s{z^2+x^2}}.
\ee
Indeed we can see that $x'z-x$ is negative because this is vanishing
at $z=0$ and (\ref{cond}) leads to $(x'z-x)'=x''z\leq 0$. In this way, we manage to
derive the g-theorem in our setup. Notice also that we can choose $x(z)$ such that $g(z)$ flows
from $g_{UV}$ to $g_{IR}$. In this case we always have $g_{UV}>g_{IR}$.

\subsection{Holographic g-theorem in Higher Dimensions}\label{gthh}

Consider a $d$ dimensional CFT on $\mathbb{R}^{1,d-1}$, whose coordinate
is denoted by $(t,x_1,\ddd,x_{d-1})$. We put a boundary at $x_1=0$ and consider
the theory on the half space defined by $x_{d-1}<0$. We assume a boundary
relevant perturbation and would like to find its holographic dual.
This gravity dual should have the translation invariance and the Lorentz invariance $SO(1,d-2)$ in the directions $(x_0,x_1,\ddd,x_{d-2})$. Therefore we can assume the following form of the metric
for the holographic dual
\be
ds^2=A(x_{d-1},z)dx_{d-1}^2+B(x_{d-1},z)dx_{d-1}dz+C(x_{d-1},z)dz^2
+D(x_{d-1},z)(-dt^2+dx_1^2+\ddd+dx_{d-2}^2). \label{solhg}
\ee
By coordinate transformations of $x_{d-1}$ and $z$, we can set $B(x_{d-1},z)=0$,
$D(x_{d-1},z)=\f{1}{z^2}$. Then by requiring the vacuum Einstein
equation (\ref{eins}), we find that allowed solutions are either pure AdS space or AdS Schwarzschild black holes. Since we are interested in zero temperature setup, the metric
should be simply given by that of the pure AdS$_{d+1}$.
We specify the boundary $Q$ by $x_{d-1}=x(z)$ again.
We can choose the null vector as follows:
\be
N^t=\pm 1,\ \ \  \ N^z=\f{1}{\s{1+(x'(z))^2}},\ \ \  \  N^{x_{d-1}}=\f{x'(z)}{\s{1+(x'(z))^2}}, \ \ \  N^{x_i}=0, \ \
(i=1,2,\ddd,d-2).
\ee
Then the condition (\ref{nulle}) leads to
\be
(K_{ab}-Kh_{ab})N^aN^b=-\f{R\cdot x''(z)}{z\left(1+(x'(z))^2\right)^{3/2}}\geq 0. \label{hgth}
\ee
As in the AdS$_3$ case, this means that $\rho_*$ is a monotonically decreasing function
under the RG flow. Therefore we can, for example, choose (\ref{gfunca}) to be an analogue of holographic g-function in higher dimension. The precise relation to the partition function
on a ball will be analyzed in the next section.

\section{AdS/BCFT in Higher Dimension}

\subsection{Holographic Dual of Balls}

Consider a $d+1$ dimensional (Euclidean) AdS space with the Poincare metric
\be
ds^2=R^2\f{dz^2+dx_\mu dx^\mu}{z^2}.
\ee
This satisfies the Einstein equation $R_{\mu\nu}-\f{1}{2}Rg_{\mu\nu}+\Lambda g_{\mu\nu}=0$.
We find
\be
R_{\mu\nu}+\f{d}{R^2}g_{\mu\nu}=0,\ \ \ R=-\f{d(d+1)}{R^2}, \ \ \  \Lambda=-\f{d(d-1)}{2R^2}. \label{eins}
\ee

To find a holographic dual of $d$ dimensional ball $B_d$ with radius $r_B$, we can act the
conformal transformation \cite{BM}
\ba
&& x'_\mu=\f{x_\mu+cx^2}{1+2(c\cdot x)+c^2\cdot x^2},\no
&& z'=\f{z}{1+2(c\cdot x)+c^2\cdot x^2},
\ea
on the $d$ dimensional half space.
In this way, we find that the following surface satisfies the constraint (\ref{eqbein})
\be
x_0^2+x_1^2+\ddd+x_{d-1}^2+\left(z-r_B\sinh(\rho_*/R)\right)^2-r_B^2\cosh^2(\rho_*/R)=0,
\ee
where $\rho_*$ is related to the tension $T$ as in (\ref{tension}). Then, we define $r(z)$ as the radius of the sphere at a slice of $z$ as follows:
\be
\label{rz def}
r(z)=\sqrt{r_B^2\cosh^2\Big(\frac{\rho_*}{R}\Big)-\Big(z-r_B\sinh \Big(\frac{\rho_*}{R}\Big)\Big)^2}.
\ee

The Euclidean action (\ref{acte}) is evaluated as follows
\begin{align}
&I_E=\dfrac{dR^{d-1}B_d}{8\pi G_N}\int^{r_Be^{\rho_*/R}}_\ep dz\dfrac{r(z)^{d}}{z^{d+1}}\ \
-\ \ \dfrac{TR^dS_d}{8\pi(d-1)G_N} \int^{r_Be^{\rho_*/R}}_{\ep}dz\dfrac{r(z)^{d-1}\sqrt{r(z)^{'2}+1}}{z^d} \nonumber \\
&=\dfrac{R^{d-1}}{8\pi G_N}\dfrac{2\pi^{d/2}}{\Gamma(d/2)}\int^{r_Be^{\rho_*/R}}_\ep dz
\left(\dfrac{r(z)^d}{z^{d+1}}-  r_B\sinh \dfrac{\rho_*}{R}\;\dfrac{r(z)^{d-2}}{z^d} \right),\quad S_d=\dfrac{2\pi^{(d+1)/2}}{\Gamma ((d+1)/2)}, \label{EUR64}
\end{align}
where $B_d$ and $S_{d}$ are the volume of $d$ dimensional unit ball and sphere, respectively.
In the last line of \eqref{EUR64}, the formula $dB_d=S_{d-1}$ and \eqref{tension} are used.
Notice that for $d=2$, the analysis has been already done in \ref{sec:be}.

\subsubsection{AdS$_5$ Case}
\label{sec ads5}

In particular we consider the AdS$_5$ case $d=4$. By subtracting the divergences simply,\footnote{
We just subtract terms of the form $\sum_{i=1}^4a_i\ep^{-i}+a_0\log\epsilon$ to make the action finite.} the result is expressed as
\be
I_E=\f{\pi R^3}{16 G_N}\left(5-2e^{-\f{2\rho_*}{R}}+4\f{\rho_*}{R}+4\log r_B\right).
\ee
For a 4D CFT with the central charge $a$, this can be rewritten as follows
\be
I_E=\f{a}{2}\left(5-2e^{-\f{2\rho_*}{R}}+4\f{\rho_*}{R}+4\log r_B\right). \label{EUC63}
\ee
Notice that the holographic g-theorem (\ref{hgth}) argues that this increases monotonically under the RG flow as opposed to the AdS$_3$ case.

Using the Euclidean action \eqref{EUC63} and holography, we can
reproduce the central charge of the
even dimensional theory via the trace anomaly. Since we consider the theory on the $d=4$ ball (a disk for $d=2$) for the field theory side, we have the scale $r_B$, namely the radius of the ball. The conformal invariance is quantum mechanically broken because of the trace anomaly. Under the Weyl transformation $\delta g_{\mu\nu}=2\delta r_Bg_{\mu\nu}/r_B$, the trace anomaly for CFT's is then described by~\cite{RT,MS}
\begin{align}
&r_B\dfrac{\delta \log Z(B_d)}{\delta r_B}=\int _{B_d}d^dx\sqrt{g}\la T^{\mu}{}_{\mu} \lb
=-2(-1)^{d/2}a\int_{B_d}\sqrt{g}E_d\nonumber \\&=-2(-1)^{d/2}A, \label{TRA64}
\end{align}
where $Z (B_d)=e^{-I_E(B_d)}$ is the partition function for the CFT,  $A$ is the central charge
in particular $A=\f{c}{12}$ for $d=2$ and $A=a$ for $d=4$. The energy stress tensor
is normalized such that $T_{\mu\nu}\equiv -(2/\sqrt{g})\delta I/\delta g^{\mu\nu}$ for the CFT's action $I$. Here, we used the fact that for theories on the ball, the following normalized Euler density $E_d$ in $d$ dimension only contributes to the trace anomaly:
\be
\int_{S_d}d^dx\sqrt{g}E_d=2,\quad \int_{B_d}d^dx\sqrt{g}E_d=1.
\ee
 Substituting the Euclidean action $I_E$
 \eqref{EUC63} or \eqref{diskpar} into $-\log Z(B_d)$ in
 \eqref{TRA64}, we indeed obtain via holography
 \be
A= \begin{cases}& a \quad \text{for $d=4$}, \\ & \frac{R}{8G_N}=\f{c}{12} \quad \text{for $d=2$}. \end{cases} \label{Cen66}
 \ee
Hence, \eqref{Cen66} reproduces the correct central charges in two and four dimensions.

For even $d=2k$ ($k\geqslant 1$), we can write the action (\ref{EUR64}) more explicitly.
As for the contribution from the bulk term, the indefinite integral we need to compute reads
\begin{equation}
\label{bulk int}
\int
\dfrac{r(z)^{2k}}{z^{2k+1}} \,dz \,=\,
(-1)^k \log z
+\sum_{r=0}^{k-1} \sum_{s=0}^{k-r}
a_{r,s}^{(k)}( \rho_\ast/R )
\left(\frac{r_B}{z}\right)^{2k-2r-s}
\end{equation}
where we have introduced
\begin{equation}
a_{r,s}^{(k)}(\alpha)\,\equiv\,
(-1)^r \binom{k}{r}\binom{k-r}{s} \frac{(2\sinh\alpha)^s}{2r+s-2k}\,.
\end{equation}
The result (\ref{bulk int}) has been obtained by expanding the polynomial $r(z)^2$ (see (\ref{rz def})).
Since  $0 \leqslant 2r+s \leqslant 2k-1$, all the terms in the double sum are power like divergent when $z \rightarrow 0$.
As for the boundary term we have to consider (now $k\geqslant 0$)
\begin{equation}
\label{bdy int}
\int \dfrac{r(z)^{2k}}{z^{2k+2}} \,dz \,=\, \frac{(-1)^{k+1}}{z}
+\frac{1}{r_B} \sum_{r=0}^{k-1} \sum_{s=0}^{k-r}
\tilde{a}_{r,s}^{(k)}(\rho_\ast/R )
\left(\frac{r_B}{z}\right)^{2k+1-2r-s}
\end{equation}
where the expansion coefficients read
\begin{equation}
\tilde{a}_{r,s}^{(k)}(\alpha)\,\equiv\,(-1)^r \binom{k}{r}\binom{k-r}{s} \frac{(2\sinh\alpha)^s}{2r+s-2k-1}\,.
\end{equation}
In the double sum occurring in (\ref{bdy int}) we have all the powers from $z^{-2k-1}$ to $z^{-2}$.\\
By employing (\ref{bulk int}) and (\ref{bdy int}) into (\ref{EUR64}), we find that all the terms computed at $z=\epsilon$ are divergent and therefore must be subtracted.
Thus $O(1)$ term in the $\epsilon$ expansion of $I_E$ is given only by the contribution at the upper extremum and it reads
\begin{eqnarray}
I_E
&=&
\frac{R^{2k-1}}{8\pi G_N}\;\frac{2\pi^{k}}{\Gamma(k)}
\bigg[ \,(-1)^k \log\big(r_B\, e^{\rho_\ast/R}\big)
+ \sum_{r=0}^{k-1} \sum_{s=0}^{k-r} a_{r,s}^{(k)}(\rho_\ast/R )\,e^{-(2k-2r-s)\rho_\ast/R} \\
\rule{0pt}{.7cm}
& & \hspace{2.4cm}
-  \;\sinh (\rho_\ast/R)
\bigg( (-1)^k e^{-\rho_\ast/R} +  \sum_{r=0}^{k-2} \;\sum_{s=0}^{k-1-r} \tilde{a}_{r,s}^{(k-1)}(\rho_\ast/R )\,e^{-(2k-1-2r-s)\rho_\ast/R}  \bigg)
\bigg].\nonumber
\end{eqnarray}
Notice also that, since the upper extremum in (\ref{EUR64}) is proportional to $r_B$, the dependence on $r_B$ cancels in all the power-like terms and enters only through the logarithmic term occurring in (\ref{bulk int}) computed at the upper extremum. This is consistent
with the CFT dual as the violation of conformal symmetry only comes from the conformal
anomaly.

\subsubsection{AdS$_4$ case}

%In this case, the ball partition function is given by
%\ba
%I_E=\f{R^2}{2G_N}\int^{r_Be^{\rho_*/R}}_\ep dz \left(\f{r^3}{z^4}-r_B\sinh\f{\rho_*}{R}\cdot\f{r}{z^3}\right).
%\ea
In this case, by evaluating $I_E$ and subtracting any divergences which are polynomials of $\ep$, we finally obtain \ba
I_E&=&\f{R^2}{2G_N}\Biggl[\f{\pi}{2}+\arctan\left(\sinh\f{\rho_*}{R}\right)
-\f{1}{24}\sinh\f{3\rho_*}{R}\no
&&\  +\left(\log \f{\ep}{r_B}+\log\cosh\f{\rho_*}{R}-\f{33}{24}-\log 2\right)\sinh\f{\rho_*}{R}\Biggr].
\ea

Notice that there is a logarithmic term which should be interpreted as some conformal anomaly.
Even though this theory is dual to three dimensional CFT, this appearance of the anomaly is expected because there is the two dimensional boundary which can lead to the conformal anomaly. Indeed the
coefficient of the log term is proportional to $\sinh\f{\rho_*}{R}$ and therefore it vanishes for
the trivial boundary $\rho_*=0$. We can define the effective boundary central charge $c_{bdy}$ as
\be
r_B\f{\de \log Z}{\de r_B}=-\f{1}{2\pi}\left\la \int dx^2\s{g}T^\mu_\mu\right\lb=\f{c_{bdy}}{6}\chi(\Sigma),
\ee
where $\chi(\Sigma)$ is the Euler number of
$\Sigma$, which is the boundary of $M$ i.e. given by $S^2$. In this way, we obtain
\be
c_{bdy}=\f{3R^2}{2G_N}\sinh\f{\rho_*}{R}.
\ee
According to the discussion in the analysis in section \ref{gthh},
$\rho_*$ decreases under
RG flows. Therefore
our boundary central charge $c_{bdy}$ also decreases under RG flows.

As done in the section \ref{sec ads5}, we can consider the generic case with even number of dimensions, i.e.
odd $d=2k-1$ ($k \geqslant 2$).
In this case the analysis is more complicated because a square root remains in the integrands of the terms involved in the computation. In order to shorten the expressions, from (\ref{rz def}) we find it convenient to introduce
\begin{equation}
\label{rtilde def}
r(z)\,\equiv \,r_B \,\tilde{r}(\tilde{z})
\hspace{2cm} \tilde{z}\,\equiv\,\frac{z}{r_B}
\end{equation}
Then, the action (\ref{EUR64}) in our case becomes
\begin{equation}
\label{action odd d}
I_E \,=\,
\frac{R^{2(k-1)}}{8\pi G_N}\;\frac{2\pi^{(2k-1)/2}}{\Gamma((2d-1)/2)}
\left[\,
\int_{\tilde{\epsilon}}^{e^{\rho_\ast/R}} \frac{\tilde{r}(\tilde{z})^{2k}}{\tilde{z}^{2k}\,\tilde{r}(\tilde{z})}\,d\tilde{z}
-\sinh\beta
\int_{\tilde{\epsilon}}^{e^{\rho_\ast/R}} \frac{\tilde{r}(\tilde{z})^{2(k-1)}}{\tilde{z}^{2(k-1)+1}\,\tilde{r}(\tilde{z})}\,d\tilde{z}
\,\right].
\end{equation}
We proceed as in the section  \ref{sec ads5} and for (\ref{rtilde def}) we get
\begin{equation}
\tilde{r}(\tilde{z})^{2h}
\,=\,
\sum_{r= 0}^h \, \sum_{s=0}^h b_{r,s}^{(h)}(\rho_\ast/R) \,\tilde{z}^{r+s}
\hspace{1.5cm}
b_{r,s}^{(h)}(\beta)  \equiv (-1)^r \binom{h}{r} \binom{h}{s} \,e^{\beta(r-s)}\,.
\end{equation}
This allows us to write the term contained between the square brackets in (\ref{action odd d}) as follows
\begin{equation}
\label{integ term odd d}
\sum_{r= 0}^k \, \sum_{s=0}^k b_{r,s}^{(k)}(\rho_\ast/R)
\int_{\tilde{\epsilon}}^{e^{\rho_\ast/R}}
\frac{d\tilde{z}}{\tilde{z}^{2k-r-s}\,\tilde{r}(\tilde{z})}\,
-\,\sinh(\rho_\ast/R)
\sum_{r= 0}^{k-1} \, \sum_{s=0}^{k-1} b_{r,s}^{(k-1)}(\rho_\ast/R)
\int_{\tilde{\epsilon}}^{e^{\rho_\ast/R}}
\frac{d\tilde{z}}{\tilde{z}^{2k-1-r-s}\,\tilde{r}(\tilde{z})}
\end{equation}
where we remark that, in the first double sum, for the power of $z^{-1}$ we have $0 \leqslant 2k-r-s \leqslant 2k$, while in the second double sum, coming from the boundary term, we have $1 \leqslant 2k-1-r-s \leqslant 2k-1$.\\
As for the integrals occurring in (\ref{integ term odd d}) we have
\begin{equation}
\int_{\tilde{\epsilon}}^{e^{\rho_\ast/R}} \frac{d\tilde{z}}{\tilde{r}(\tilde{z})}
\,=\, \frac{\pi}{2}+\arctan (\sinh(\rho_\ast/R)) + \dots
\qquad
\int_{\tilde{\epsilon}}^{e^{\rho_\ast/R}}  \frac{d\tilde{z}}{\tilde{z}\,\tilde{r}(\tilde{z})}
\,=\, \log r_B -\log\left(\frac{\cosh(\rho_\ast/R)}{2}\right)  + \dots
\end{equation}
By computing other examples, we recognize the following structure for $q \geqslant 2$
\begin{eqnarray}
\int_{\tilde{\epsilon}}^{e^{\rho_\ast/R}}  \frac{d\tilde{z}}{\tilde{z}^q\,\tilde{r}(\tilde{z})}
&=&
\bigg[ \,C_q(\rho_\ast/R)\, \log\left(\frac{\tilde{z}}{1+\tilde{z} \sinh(\rho_\ast/R)+\tilde{r}(\tilde{z})}\right)
-P_{q-2}(\tilde{z},\rho_\ast/R)\,\frac{\tilde{r}(\tilde{z})}{\tilde{z}^{q-1}}\,
\bigg] \bigg|_{\tilde{\epsilon}}^{e^{\rho_\ast/R}} \nonumber\\
\rule{0pt}{.8cm}
&=&  C_q(\rho_\ast/R)  \left[\,\log r_B -\log\left(\frac{\cosh (\rho_\ast/R)}{2}\right) \right]   + \gamma_q(\rho_\ast/R)
+ \dots
\end{eqnarray}
where the dots denote vanishing or diverging terms when $\epsilon \rightarrow 0$.
The polynomial $P_{q-2}(\tilde{z},\beta)$ has degree $q-2$ in terms of $\tilde{z}$  and its coefficients depend on $\rho_\ast/R$. We recall  that $\tilde{r}(e^{\rho_\ast/R})=0$ and $\tilde{\epsilon}= \epsilon/r_B$.
Unfortunately we are not able to get closed forms for the function $C_q(\rho_\ast/R)$ and $P_{q-2}(\tilde{z},\rho_\ast/R)$ and, in order to obtain at least the dependence on $r_B$ of the finite part of the action for any $k$ we need to know $C_q(\rho_\ast/R)$ for any positive integer $q\geqslant 2$.

\subsection{Holographic Dual of Half Space}

Consider a holographic dual of a half space in the pure AdS$_{d+1}$
\ba
&& ds^2=R^2\f{dz^2+\sum_{i=0}^{d-2}dx_i^2+dx_{d-1}^2}{z^2}.
\ea
According to (\ref{corc}), the boundary $Q$ is defined by
\be
x=\sinh\f{\rho_*}{R}\cdot z,
\ee
where $\rho_*$ is related to the tension via $TR=(d-1)\tanh\f{\rho_*}{R}$.
The length in the Euclidean time and the one in $x_{d-1}$ direction
are defined to be $\beta$ and $L$.
The Euclidean action $I_E$ is given by
\ba
I_E&=&\f{d R^{d-1} V_{d-2}\beta L}{8\pi G_N}\int^{\infty}_\ep\f{dz}{z^{d+1}}+\f{(d-1)R^{d-1}V_{d-2}\beta}{8\pi G_N}\sinh\f{\rho_*}{R}\int^{\infty}_\ep\f{dz}{z^d}\no
 &=& \f{R^{d-1} V_{d-2}\beta L}{8\pi G_N\ep^{d}}+\f{R^{d-1}V_{d-2}\beta}{8\pi G_N\ep^{d-1}}\sinh\f{\rho_*}{R}.
\ea
Both divergent terms are completely canceled by the counter terms. Therefore $I_E=0$.

\subsection{Holographic Dual of Strips}

Consider a holographic dual of a strip in $\mathbb{R}^{1,d-1}$ which is defined by
restricting to the range $0<x_{d-1}<\Delta x$ with
other coordinates $x_0,x_1,\ddd,x_{d-2}$ taking any values. We are interested in the zero temperature case and thus we consider the AdS soliton geometry
\be
ds^2=R^2\f{h(z)^{-1}dz^2+\sum_{i=0}^{d-2}dx_i^2+h(z)dx_{d-1}^2}{z^2},
\hspace{1.6cm}
h(z)=1-\left(\f{z}{z_0}\right)^{d}.
\ee
 We compactify $x_{d-1}$ as $x_{d-1}\sim x_{d-1}+\f{4\pi}{d}z_0$.
 We describe the boundary $Q$ by $x_{d-1}=x(z)$. The constraint (\ref{eqbein}) leads to
 the differential equation
 \be
 x'(z)=\pm\f{RT}{h(z)\s{(d-1)^2h(z)-R^2T^2}},
 \ee
 where $+$ (or $-$) sign corresponds to the case where the bulk spacetime $N$ is situated
 in the side of smaller (or larger) $x$. We fix the integration constant by requiring $x(z_*)=0$,
 where $z_*$ is defined by $(d-1)^2h(z_*)=R^2T^2$.
   Notice that the bulk spacetime $N$ is given by
\be
-x(z)\leq x_{d-1} \leq x(z), \label{setr}
\ee
for $T\leq 0$. For $T>0$, $N$ is given by the complement of (\ref{setr}) at $T=-|T|$
with respect to the total manifold of the AdS soliton i.e. $|x_{d-1}|\leq \f{2\pi}{d}z_0$.

If we define $x_0\equiv x(0)$, the length $\Delta x$ of the AdS boundary $M$ is given by
\ba
\Delta x=2x_0\ \ \ (\mbox{when}\ T\leq 0),\no
\Delta x=\f{4\pi}{d}z_0-2x_0\ \ \  (\mbox{when}\ T\geq 0).
\ea

We can calculate $x_0$ as follows
\ba
x_0&=&\int^{z_*}_0 \left(\f{dx}{dz}\right)dz=\f{RTz_0}{d(d-1)}\int^{w_*}_0
dw\f{w^{\f{1-d}{d}}}{(1-w)\s{1-\f{R^2T^2}{(d-1)^2}-w}}, \no
&=& \f{\Gamma(1/d)\Gamma(1/2)}{\Gamma(1/2+1/d)}\cdot\f{RTz_0}{d(d-1)}\cdot
w_*^{1/d-1/2}\cdot F(1,1/d,1/2+1/d;w_*),
\ea
where $w_*\equiv 1-\f{R^2T^2}{(d-1)^2}$ and $F(a,b,c;z)$ is the standard hypergeometric function.

The entanglement entropy can be calculated as follows (we first assume $T<0$):
\ba
I_E&=&-\f{1}{16\pi G_N}\int_N\s{g}(R-2\Lambda)-\f{1}{8\pi G_N(d-1)}\int_Q\s{h}T,\no
&=& \f{dR^{d-1}L^{d-2}\beta}{4\pi G_N}\int^{z_*}_\ep dz\f{x(z)}{z^{d+1}}
-\f{TR^{d}L^{d-2}\beta}{4\pi G_N}\int^{z_*}_{\ep}\f{dz}{z^{d}\s{(d-1)^2h(z)-R^2T^2}},\no
&=&-\f{R^{d-1}L^{d-2}\beta}{4\pi G_N}\left[\f{h(z)x(z)}{z^d}\right]^{z_*}_\ep,\no
&=& \f{R^{d-1}L^{d-2}\beta}{4\pi G_N}x_0\cdot\left(\f{1}{\ep^d}-\f{1}{z_0^d}\right)
-\f{R^{d-1}L^{d-2}\beta }{4\pi G_N \ep^{d-1}}\cdot\f{RT}{\s{(d-1)^2-R^2T^2}}. \label{highfin}
\ea
Here $\beta$ and $L$ is the length of $x_0$ and $x_i$ directions.

We would like to cancel the divergences by adding local counter terms.
The bulk counter term which is proportional to $\ep^{-d}$ leads to the subtraction
(remember the factor $\s{h(\ep)}$ in the $x_{d-1}$ direction):
\be
-\f{R^{d-1}L^{d-2}\beta}{4\pi G_N}x_0\cdot\f{\s{h(\ep)}}{\ep^d}. \label{fince}
\ee
Also the final term in (\ref{highfin}) is simply canceled by the boundary counter term.
In this way, by taking into account the finite contribution in (\ref{fince}) we obtain the
final result
\be
I_E=-\f{R^{d-1}L^{d-2}\beta}{16\pi G_N}\cdot\f{\Delta x}{z_0^d}. \label{evnhig}
\ee
It is possible to check that this result (\ref{evnhig}) remains the same even if $T>0$.

\section{Time-dependent Configurations}

So far we have focused on the holography for static systems where the time evolution  of
boundaries in a CFT is trivial. Thus here we would like to study time dependent configurations
especially in the AdS$_3/$BCFT$_2$ setup. In most of the arguments below, higher dimensional
generalizations are straightforward.

\subsection{Description in terms of Entangled Pair}\label{entsec}

\begin{figure}[ttt]
   \begin{center}
     \includegraphics[height=5cm]{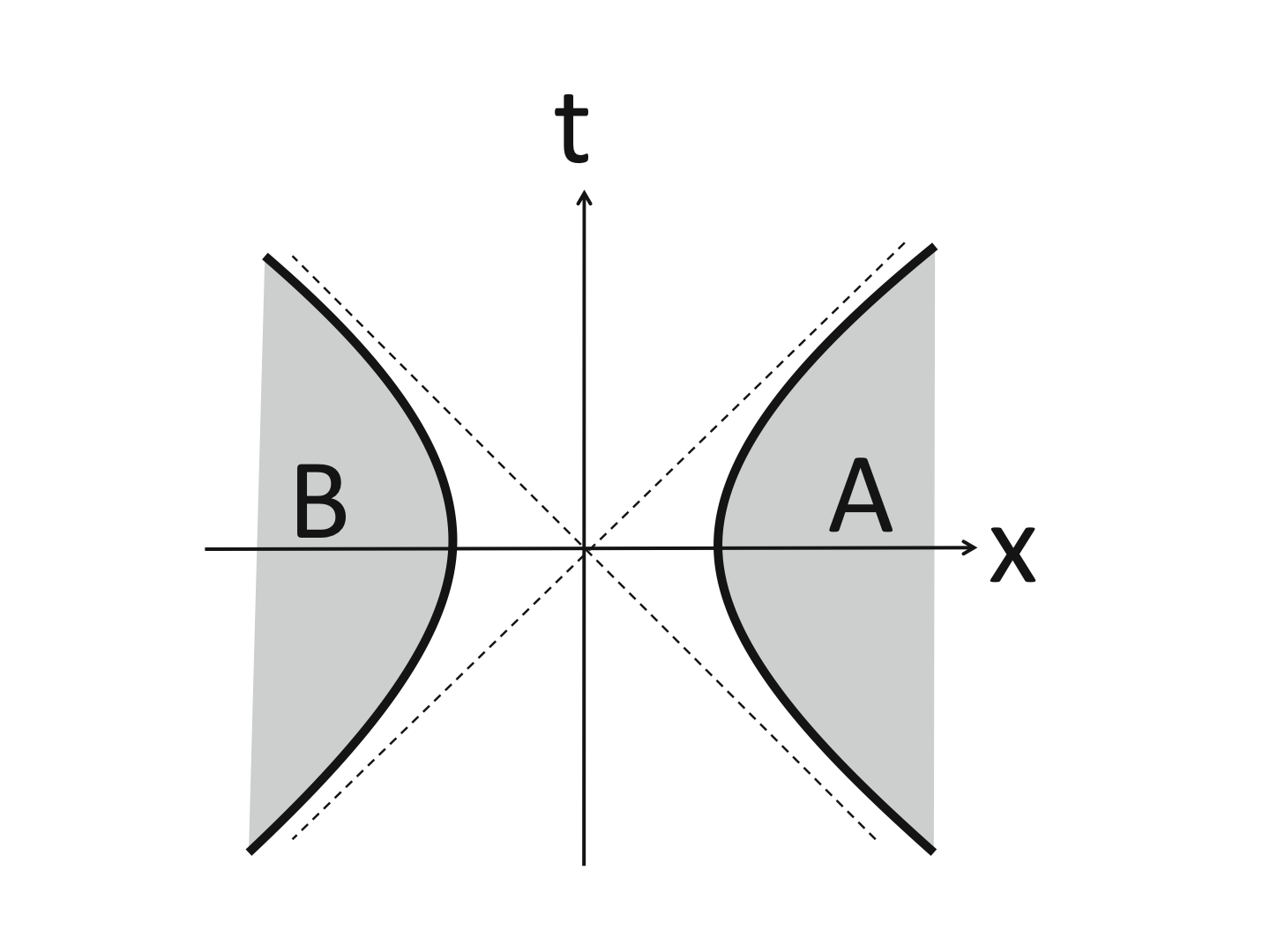}
     \hspace*{0.5cm}
       \includegraphics[height=5cm]{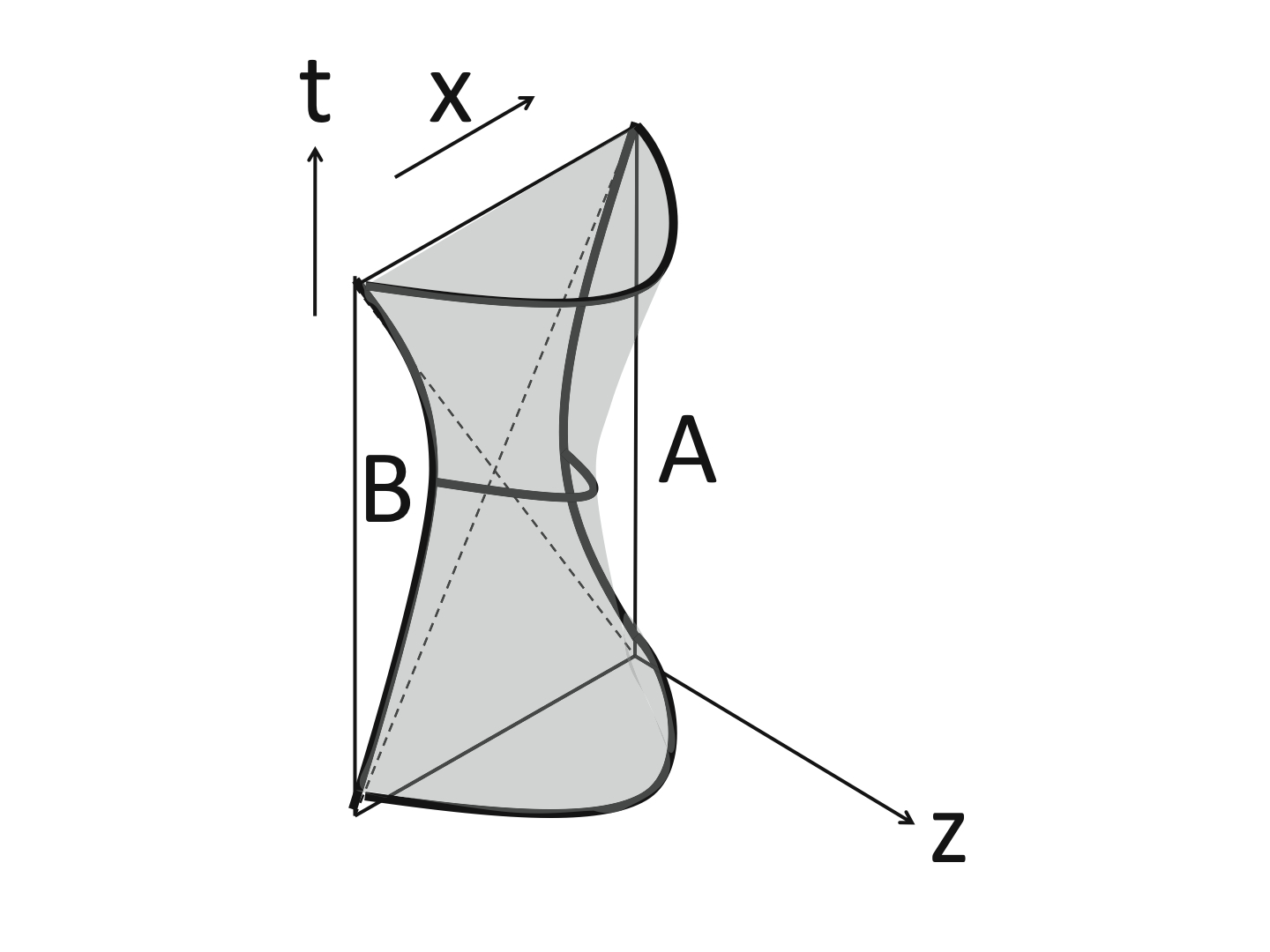}
   \end{center}
   \vspace{-.8cm}
   \caption{The left figure denotes a two dimensional CFT defined on a spacetime with time-dependent boundaries. The right figure describes its holographic dual.}\label{fig:time}
\end{figure}

Consider a pure AdS$_3$
\be
ds^2=R^2\f{-dt^2+dz^2+dx^2}{z^2}, \label{adsyt}
\ee
and a boundary $Q$ specified by $z=z(x,t)$. The solutions to the boundary equation of motion (\ref{eqbein})
are obtained from (\ref{diskh}) by a double Wick rotation. The gravity dual is given by the region \be
(z-A)^2+(x-\ap)^2-(t-\beta)^2\ge \gamma^2, \label{entone}
\ee
where $\ap,\beta,\gamma$ and $A$ are arbitrary constants. This is depicted in
the right figure in Fig.\ref{fig:time}. The tension $T$ is related to
these parameters via $T=-\f{A}{R|\gamma|}$. We also need to assume $|A|<|\gamma|$ so that the
boundary of (\ref{entone}) is time-like. Using the translational invariance we simply set $\ap=\beta=0$ below. The AdS boundary of this spacetime (\ref{entone}) consists of two disconnected regions $x\ge \s{\gamma^2-A^2+t^2}$ and $x\le -\s{\gamma^2-A^2+t^2}$, which are connected in the
bulk of AdS (see the left figure in Fig.\ref{fig:time}).

It is also possible to flip the sign of the inequality in (\ref{entone}). This leads to
the region $(z-A)^2+(x-\ap)^2-(t-\beta)^2\le \gamma^2$. In this case, the region looks like a half of three dimensional cylinder and this is dual to a CFT lives on an interval defined by
$|x|\leq \s{\gamma^2-A^2+t^2}$.

Let us consider the holographic interpretation of the solution (\ref{entone}).
Since end point of each of boundaries is accelerated much like a
Rindler observer, the light cone defined by $t=\pm x$ plays the role of Rindler horizon.
Therefore the two regions at the AdS boundary are causally disconnected. However, they are
connected in the bulk AdS. This means that they are entangled with each other as in the
AdS Schwarzschild black holes \cite{MaBH}. This interpretation gets clearer if we
perform the following familiar coordinate transformation:
\ba
&& t-x=-e^{\theta-u}\s{1-\f{r_+^2}{r^2}},\no
&& t+x=e^{\theta+u}\s{1-\f{r_+^2}{r^2}},\no
&& z=\f{r_+e^{\theta}}{r}.  \label{cortf}
\ea
This leads to the BTZ black hole metric
\be
ds^2=R^2\left[-\left(\f{r^2}{r_+^2}-1\right)du^2+\f{dr^2}{r^2-r_+^2}+\f{r^2}{r_+^2}\,d\theta^2\right],
\label{bhd}
\ee
where the horizon is situated at $r=r_+$. In this coordinate system, (\ref{entone}) can be
rewritten as
\be
e^{2\theta}-2\f{Ar_+e^{\theta}}{r}\ge \gamma^2-A^2, \label{constrs}
\ee
which coincides with (\ref{solx}).~\footnote{See also appendix in which we also get a surface in the rotating BTZ BH using the coordinate transformation.} The region $|x|\ge \s{\gamma^2-A^2+t^2}$, where the CFT is defined, is now mapped to the half line $\theta\ge \f{1}{2}\log(\gamma^2-A^2)$. The horizon
$r=r_+$ is mapped to the Rindler horizon in (\ref{adsyt}).

The holographic entanglement entropy between the CFT on the two half lines
(we choose the subsystem $A$ as one of the half lines) are computed as follows\footnote{
Strictly speaking, we need to apply the covariant prescription of holographic entanglement entropy
\cite{cov} in such a Lorentzian time-dependent spacetime. However, since here we restrict to the time symmetric surface at $t=0$, we just need to calculate the geodesic length on that surface.}
\be
S_A=\f{R}{4G_N}\int^{z_{IR}}_{\gamma+A}\f{dz}{z},
\ee
where $z_{IR}$ is the IR cut off, which is very large. We can easily see that this is equivalent to the entropy of the BTZ black hole (\ref{bhd}) restricted inside of the boundary $Q$. Since the condition (\ref{constrs}) is estimated as $e^\theta\ge A+\gamma$, we find the black hole entropy
is given by
\be
S_{BH}=\f{R}{4G_N}\int^{\log z_{IR}}_{\log(\gamma+A)} d\theta.
\ee
Therefore $S_A$ and $S_{BH}$ are identical as we wanted to show.

These correspondences are naturally understood as an example of
Unruh effect because a CFT on two half lines which are entangled due to a uniform acceleration, is equivalent to a CFT on a half line at finite temperature.

It is straightforward to generalize our setups so that it is dual to a CFT on
the two disconnected intervals, instead of half lines. This is obtained by considering the gravity on the region
\be
\gamma^2_1\leq (z-A)^2+x^2-t^2\leq \gamma^2_2, \label{rfgk}
\ee
assuming $|\gamma_1|<|\gamma_2|$. After the coordinate transformation (\ref{cortf}), the region
(\ref{rfgk}) is mapped to an interval in the BTZ black hole.

\subsection{Comments on g-theorem and Topological Censorship}

In the standard AdS/CFT, a Lorentzian asymptotic AdS spacetime with multiple
AdS boundaries which are connected without encountering any horizons,
is prohibited once we assume an appropriate energy condition \cite{
Worm}. This is called the topological censorship. It is interesting to study
a similar property in our AdS/BCFT setups.
We only consider static setups and assume that the AdS boundary consists of two
disconnected intervals.

In the presence of two intervals, we can find a wormhole like geometry if we can connect
them in the bulk AdS. However, this is not allowed once the null energy condition is imposed.
As discussed in section \ref{gthh}, the g-theorem is holographically equivalent to $x''(z)\leq 0$.
This requires the boundary $Q$ eventually evolves toward the inside direction. Therefore it prevents the boundaries from connecting smoothly with each other. In this way, we find that the
g-theorem is holographically interpreted as the topological censorship. Notice that in our previous example of section \ref{entsec}, the two disconnected regions at the AdS boundary can connect inside of the bulk AdS because the two regions are connected through a horizon.

\section{Correlation Functions}
\label{corr section}

\subsection{One Point Function}

Considering the setup given by the metric (\ref{metads}), we employ the
metric (\ref{poin}).  The boundary $Q$ is specified by \be x\,=\,\tan
\theta \,z, \ee where $\tan\theta=\sinh\f{\rho_*}{R}$ (see (\ref{corc})).

We are interested in a one point function $\la O\lb$ of a scalar
operator $O$ and the scalar field dual to $O$ is denoted by $\phi$.
The bulk action of $\phi$ is given by
\ba
S&=&\dfrac{1}{2}\int (dx)^{d+1}\s{g}
(\de^\mu\phi\de_\mu\phi+m^2\phi^2)\no &=& \dfrac{1}{2}\int
dz\, dx (dw)^{d-1}
\left[\f{R^{d-1}}{z^{d-1}}\Big((\de_z\phi)^2+(\de_x\phi)^2+(\de_\tau\phi)^2+(\de_{\vec{w}}\phi)^2\Big)
+\f{R^{d+1}}{z^{d+1}}m^2\phi^2 \right].
\label{busc}
\ea
where $(dw)^{d-1} = d\tau d\vec{w}$.
On the boundary $Q$, we assume the following coupling of $\phi$
\be
S_Q=\int_{Q} (d\xi)^d \s{h}(a\phi+\ddd),
\ee
where the coordinates $\xi$ parameterize the boundary $Q$ and $\ddd$ means higher
terms in $\phi$ which are not considered here; $a$ is just a constant.

It is useful to define a rotated coordinate $\ti{x}$ and $\ti{z}$
\be \ti{x}=\cos\theta x-\sin\theta z,\ \ \ \ \ti{z}=\sin\theta
x+\cos\theta z. \ee The boundary $Q$ is described by $\ti{x}=0$. By
considering the variation $\delta \phi$ of $\phi$ with the equation
of motion imposed, we find \be \delta S=R^{d-1}\int
d\ti{z}(dw)^{d-1}\delta
\phi\left(\f{\de_{\ti{x}}\phi}{z^{d-1}}+\f{aR}{z^d}\right). \ee

In this way, we find that the boundary condition on $Q$ for the scalar
field $\phi$ reads \be (\cos\theta
\de_x-\sin\theta\de_z)\phi+\f{aR}{z}=0, \label{bcsc} \ee
where the operator acting on $\phi$ is $\partial_{\tilde{x}}\;$.
Notice that
to calculate the one point function, we can neglect the dependence
on $w$. \\
Requiring the regularity at $z=\infty$, we can expand the
solution to the equation of motion of (\ref{busc}) as follows \be
\phi(z,x)=z^{d/2}\int ^{\infty}_{-\infty}dk A(k)K_\nu(|k|z)e^{ikx}, \label{Phi7}
\ee where $\nu=\s{d^2/4+m^2R^2}$.
 The function $A(k)$ can be found by plugging (\ref{Phi7}) into the boundary condition (\ref{bcsc}) which becomes
\begin{align}
\label{bcsol}
& \int_{-\infty}^{+\infty}A(k) \left[\,
ik\,K_\nu(|k| z) \cos\theta-\left(\frac{d}{2z}\,K_\nu(|k| z)+\partial_z K_\nu(|k| z)\right)\sin\theta\,
\right] e^{ikz \tan\theta} dk\\
&=\int_{-\infty}^{\infty} dk \left[ikA(k)\cos\theta K_\nu(|k|z)e^{ikz\tan\theta}-
 \dfrac{\sin\theta}{z} \Big(\dfrac{dA(k)e^{ikz\tan\theta}}{2}-\Big(kA(k)e^{ikz\tan\theta}\Big)'\Big)K_\nu(|k|z)\right] \nonumber \\
&=-\f{aR}{z^{d/2+1}}, \nonumber
 \end{align}
where in the last line, we performed a partial integration in terms of $k$ using $\de_z K_{\nu}(|k|z)=k\de_k K_{\nu}(|k|z)/z$. In the appendix \ref{apsf} we give a detailed treatment of this equation.

By taking the AdS boundary limit $z\to 0$ of $\phi$, we get the well known
behavior
\be
\phi\to z^{d-\Delta}\ap(x)+z^{\Delta}\beta(x), \label{bsc9}
\ee
where $\Delta=d/2+\nu$ is the conformal dimension of the dual
operator. According to the standard dictionary in AdS/CFT~\cite{KW}, we find
the one point function as \be \la O\lb = (2\Delta -d)\beta(x). \label{One10} \ee
Note that
 we have imposed two boundary conditions to determine the behavior of the solution \eqref{Phi7}. The boundary condition of the regularity at $z=\infty$ is included in the boundary condition~\eqref{bcsc} and the boundary condition at the AdS boundary limit $z\to\infty$ is also imposed on $\phi$ as seen in \eqref{bsc9}. Thus, the boundary condition above is enough to determine the solution of $\phi$.
 The dependence on $k$ of $A(k)$ turns out to be\,\footnote{If we choose the ansatz for the function $A(k)\sim |k|^{d/2-1}$, the left hand side of the boundary condition \eqref{bcsol} vanishes.}
\be
\label{cdef}
A(k)\,\equiv\, c_\theta\,\frac{|k|^{d/2}}{k}
\ee
and in the following we will find $c_\theta$.
In the simplest case of $\theta=0$, from (\ref{bcsol}) we have
\be
\label{Ak0}
c_0 \,=\,
\dfrac{iaR}{2^{d/2}\Gamma \Big(\frac{d-\Delta +1}{2}\Big)\Gamma \Big(\frac{\Delta+1}{2}\Big)}
\hspace{1.4cm}\text{for $\;d/2+1>\nu$}
\ee

\begin{figure}[tb]
   \begin{center}
     \includegraphics[height=8cm]{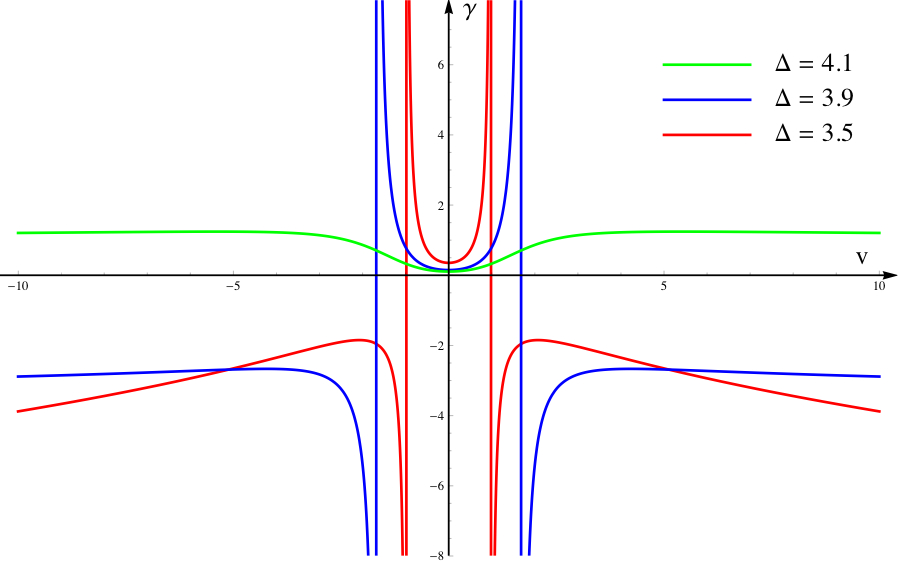}
   \end{center}
      \vspace{-.4cm}
   \caption{Setting $R=1$, $v(=\sinh \rho_*/R)$ dependence of $\gamma$ for $d=4$ is plotted, where $\gamma$ is the coefficient of the one-point function. $|\gamma| $ for the irrelevant operator ($\Delta >4$) seems to be small for any $v$. }\label{fig:1point}
\end{figure}
\noindent (see the appendix \ref{apsf} for details). Thus we find
 \begin{eqnarray}
\beta(x)\;=\;\dfrac{\la {O} \lb}{2\Delta -d}
&=&
c_0\int_{-\infty}^{\infty} dk \dfrac{-\pi k^{-1}|k|^{\Delta}}{2\sin\pi \Big(\Delta -\frac{d}{2}\Big)\Gamma \Big(\Delta -\frac{d}{2}+1\Big)2^{\Delta -d/2}}e^{ikx} \nonumber \\
\rule{0pt}{.9cm}
&=&\dfrac{\pi aR\sin \Big(\frac{\pi\Delta}{2} \Big)\Gamma (\Delta)\;x^{-\Delta}}{2^{\Delta }\sin\pi \Big(\Delta -\frac{d}{2}\Big)\Gamma \Big(\Delta -\frac{d}{2}+1\Big)\Gamma \Big(\frac{d-\Delta +1}{2}\Big) \Gamma \Big(\frac{\Delta+1}{2}\Big)}. \label{POI12}
\end{eqnarray}
In this way we get $\la O\lb \sim 1/x^{\Delta}$ as expected. Our result can be extended for $\theta\neq 0$.  Assuming (\ref{cdef}), the boundary condition \eqref{bcsol} is solved by
\be
\label{ctheta}
c_\theta\,=\,\dfrac{iaR}{2^{d/2}\Gamma \Big(\frac{\Delta +1}{2}\Big)\Gamma \Big(\frac{d-\Delta +1}{2}\Big)\sqrt{1+v^2}F(\frac{\Delta +1}{2},\frac{d-\Delta +1}{2},\frac{1}{2};-v^2)}
\ee
(again, see the appendix \ref{apsf} for details) where $v=\sinh (\rho_*/R)$ and therefore we get
\begin{eqnarray}
\la O \lb \;\equiv\; \frac{aR\gamma}{x^{\Delta}}
&=&
\dfrac{ aR(2\Delta -d)\pi\sin \Big(\frac{\pi\Delta}{2} \Big)\Gamma (\Delta)}{2^{\Delta }\sin\pi \Big(\Delta -\frac{d}{2}\Big)\Gamma \Big(\Delta -\frac{d}{2}+1\Big)\Gamma \Big(\frac{d-\Delta +1}{2}\Big) \Gamma \Big(\frac{\Delta+1}{2}\Big)}
\nonumber \\
\rule{0pt}{.8cm}
& & \cdot\;\dfrac{x^{-\Delta}}{\sqrt{1+v^2}F(\frac{\Delta +1}{2},\frac{d-\Delta +1}{2},\frac{1}{2};-v^2)}\; .
\end{eqnarray}
Thus, we find that the scaling of the one-point function is independent of $\theta$. We can also see from
  above that the coefficient of the one-point function should be real by considering the odd/even property of the Fourier integral \eqref{bcsol}.
 We plotted the coefficient of the one-point function $\gamma $ as the function of $v$ in Fig. \ref{fig:1point}.

For $d=4$, it is interesting to compare the one-point function for $\theta=0$ \eqref{One10} with the one-point function obtained via the AdS/dCFT correspondence~\cite{DeWolfe:2001pq}.
Substituting $aR=\lambda^{1/2}$ into \eqref{POI12} and using some formulas for the Gamma function, we can confirm that our one-point function agrees completely with the one-point function in~\cite{DeWolfe:2001pq} of the following form:
\be
\la O\lb=\lambda^{1/2}
\dfrac{\Gamma \Big(\frac{\Delta -3}{2}\Big)\Gamma \Big(\frac{\Delta}{2}\Big)\Gamma \Big(\frac{3}{2}\Big)}{\pi \Gamma(\Delta -2)}\, x^{-\Delta}.
\label{ONP111}
\ee
As pointed out in~\cite{DeWolfe:2001pq}, the power of $\lambda$ in \eqref{ONP111} at strong coupling does not need to agree with field theory results at weak coupling. For general BCFT with the gravity dual, this power at strong coupling seems to be related with the parameter $aR$.

\section{An Example of String Theory Embedding}

As a final study in this paper, we would like to present a concrete setup in string theory which can be regarded as
an example of AdS/BCFT. In our model described below, orientifolds with D-branes play the role of the boundary $Q$. In a geometry which looks like $AdS\times M$ ($M$ is a compact manifold), we would like to wrap
the orientifolds and branes completely on $M$ so that the backreaction inside $M$ gets trivial and
that only the AdS part becomes relevant as assumed in our effective descriptions of AdS/BCFT.

We would like to consider $AdS_4\times CP^3$ solution in type IIA string theory
dual to $N=6$ supersymmetric Chern-Simons theory (ABJM theory) \cite{ABJM}.
We write the metric of $AdS_4$ as follows
\be
ds^2=R^2\f{dz^2-dt^2+dx^2+dy^2}{z^2}.
\ee
The ABJM theory lives on $(t,x,y)$ and consists of the $U(N)_k\times U(N)_{-k}$ gauge field, denoted by $(A^{(1)}_\mu,A^{(2)}_\mu)$ and four chiral multiplets (its bi-fundamental
scalars and fermions are
denoted by $X^I$ and $\Psi^I$ ($I=1,2,3,4)$. We introduce the two orientifold 8-planes ($O8$-planes) so that they are situated at $y=0$ and $y=L$, extending in the nine directions other than $y$.
Notice that all of such configurations preserve the
$SU(4)$ R-symmetry. There are four different types of
$O8$-planes denoted by $O8^{-}$, $\ov{O8}^{\ -}$, $O8^{+}$ and $\ov{O8}^{+}$
(for such a classification of orientifolds see e.g.\cite{WiB,ABJ,Ho}). Their
differences are clear in the boundary state formalism. If we write the cross-cap state in the NSNS and RR sector as $|\Omega_{NSNS}\lb$ and $|\Omega_{RR}\lb$, then we can represent
\ba
&& |O8^{-}\lb=-|\Omega_{NSNS}\lb-|\Omega_{RR}\lb,\ \
|\ov{O8}^{\ -}\lb=-|\Omega_{NSNS}\lb+|\Omega_{RR}\lb ,\no
&& |O8^{+}\lb=|\Omega_{NSNS}\lb+|\Omega_{RR}\lb,\ \
|\ov{O8}^{+}\lb=|\Omega_{NSNS}\lb-|\Omega_{RR}\lb.
\ea
In string theory setup with orientifolds, we need to impose the tap-pole cancelation, which
requires that the total RR charge vanishes. For example, the RR-charge of an $O8^{\mp}$-plane is given by $\mp 8$ times that of a D8-brane.

\begin{figure}[ttt]
   \begin{center}
     \includegraphics[height=5cm]{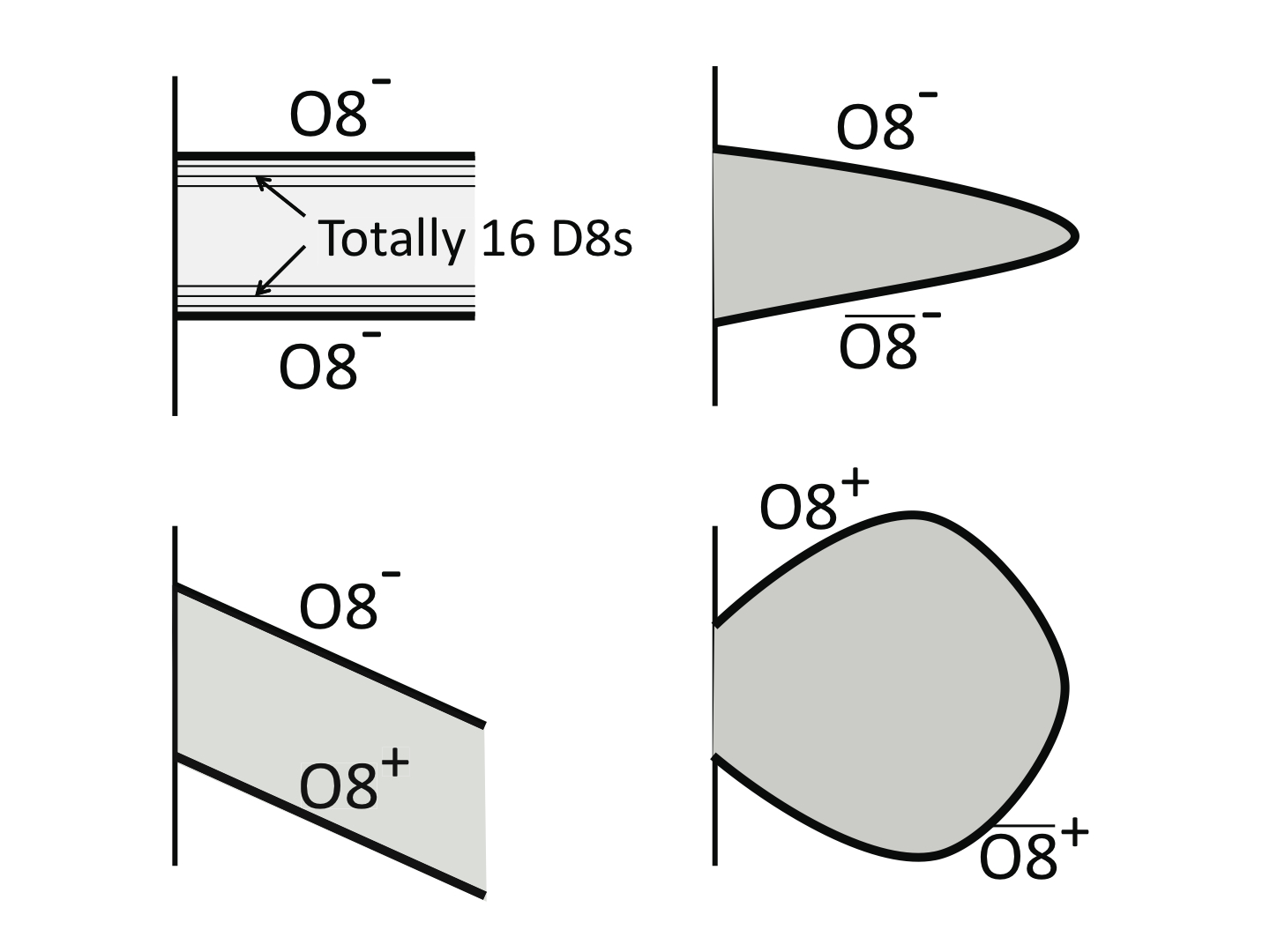}
   \end{center}
      \vspace{-.6cm}
   \caption{Schematic Pictures of AdS$_4/$BCFT$_3$ setups with various O8-planes. The vertical line in the left side in each of figures represents the AdS boundary.}\label{fig:orientifolds}
\end{figure}

The basic setup is the $O8^{-}-O8^{-}$ system, i.e. a $O8^{-}$ is at $y=0$ and another $O8^{-}$ is
at $y=L$ (refer to Fig.\ref{fig:orientifolds}). We need to add totally 16 D8-branes. If we place a half of them at $y=0$ and the other half at $y=L$, then the RR charge source completely cancels out and the solution will coincide with the one in AdS/BCFT with the vanishing tension $T=0$ as the NS tadpole is also completely canceled out. In this case of two $O8^{-}$ planes, the $Z_2$ orientifold projection acts as a parity transformation which exchanges the two gauge groups
\ba
&& A^{(1,2)}_{t,x}(t,x,y)\to -\left(A^{(2,1)}_{t,x}(t,x,-y)\right)^T,\no
&& A^{(1,2)}_{y}(t,x,y)\to \left(A^{(2,1)}_{y}(t,x,-y)\right)^T,\no
&& X^I(t,x,y)\to X^I(t,x,-y)^{T},\no
&& \Psi^I(t,x,y)\to P\cdot\Psi^I(t,x,-y)^{T},\label{proj}
\ea
where $M^T$ denotes the matrix transposition of $M$; the matrix $P$ is the standard one which describes the parity transformation of
three dimensional fermions. For more details refer to the appendix \ref{discs}. Notice that
the $SU(4)_R$ symmetry which rotates the index $I$ is preserved after this projection.

The $Z_2$ projection (\ref{proj}) is also clear from the T-dualized setup of $N$ D3-branes
with a NS5-brane and $(1,k)$ 5-brane considered in \cite{ABJM}. The D3-branes are wrapped on
a circle and they are divided into two parts (called fractional D3-branes), cut by the two 5-branes. We insert two O7-planes so that they coincide with two 5-branes. The orientifold action now exchanges two different fractional D3-branes and therefore it also
exchanges the two $U(N)$ gauge group. This setup preserves a half of the original supersymmetries.

The presence of 8 D8-branes at each of the boundaries $y=0$ and $y=L$ introduces 8 massless chiral (complex) fermions at each of them. This comes from the D2-D8 open strings with the orientifold
projection. In the ABJM setup without the orientifold, an appearance of Weyl fermions have been studied in \cite{FLRT} as an example of holographic edge states of quantum Hall effect.
In summary, this three dimensional gauge theory dual to the AdS$_4$ with two $O8^{-}$ planes is defined by the ABJM theory projected by the action (\ref{proj}) at two boundaries $y=0$ and $y=L$ with 8 chiral fermions at each of the boundaries. This supersymmetric Chern-Simons theory lives on $R^{1,2}\times S^1/Z_2$ and note that $S^1/Z_2$ describes an interval.
The holography tells us that this theory preserves the boundary conformal symmetry at the two boundaries and therefore it offers us an example of AdS/BCFT. The tension of the boundaries is vanishing as the
local tadpole cancelation between O8-planes and D8-branes.

Before we go on, it is instructive to why these edge modes need to appear in this model. To answer this, we would like to look at the gauge anomaly when we reduce the theory on the interval and
consider its two dimensional field theory in the low energy limit. It is useful to note that the matrix $P$ can be regarded as the chirality matrix when it is reduced to two dimension.
Thus the condition (\ref{proj}) tells us that the right-moving mode of the fermion should
be symmetric, while the left-moving one should be antisymmetric. In this way, the low energy
two dimensional theory is not a non-chiral theory and we have to worry about the gauge anomaly.
Now we want to
estimate the anomaly of $U(N)$ gauge symmetry which is obtained from the projection of $U(N)\times U(N)$ gauge symmetry by (\ref{proj}). First consider the $SU(N)$ part. The gauge anomaly induced matter in the representation $R$ in two dimension is proportional to
Tr$_R[T^aT^b]_{right}-$Tr$_R[T^aT^b]_{left}$, where the difference means that between the left-moving sector contribution and right-moving one. The relation of such traces between antisymmetric
(A), symmetric (S) and fundamental representation (F) is given by
\be
\mbox{Tr}_{A}[T^aT^b]=(N-2)\mbox{Tr}_F[T^aT^b],\ \ \ \ \mbox{Tr}_{S}[T^aT^b]=(N+2)\mbox{Tr}_F[T^aT^b].
\ee
Since there are four Dirac fermions $\Psi^I$ which satisfy the projection (\ref{proj}), the $SU(N)$ gauge anomaly is given by
\be
4\mbox{Tr}_{S}[T^aT^b]-4\mbox{Tr}_{A}[T^aT^b]=16\mbox{Tr}_F[T^aT^b].
\ee
This anomaly is completely canceled if we add extral 16 chiral left-moving fermions with the fundamental representation, which indeed
coincide with the edge modes from the D8-branes. For the anomaly for the $U(1)$ part of
$U(N)$, we can similarly confirm the cancelation.

Moreover, we can move some of the D8-branes from one of $O8^{-}$ planes to another with the supersymmetries
kept. This offers us an example where the boundary $Q$ has non-vanishing tension. However, we need to be careful to identify its bulk geometry because the boundary also gives rise to the dilaton
gradient. It will be an intriguing future problem to work out its back-reacted solution in type IIA supergravity.

So far we studied the case where both of the O8-planes are the $O8^{-}$ type. We would also like to briefly discuss some other cases (refer to Fig.\ref{fig:orientifolds}).
If one of them is $O8^{-}$ and the other is its anti-plane i.e. $\ov{O8}^{-}$, then the supersymmetries are completely broken and we do not need any D8-branes for the RR tadpole cancelation. In this case, no chiral fermion appears and there is no anomaly.
This is also consistent with the fact that the fermion boundary condition is twisted and no zero mode remains. Notice that the $Z_2$ projection for $\ov{O8}^{-}$
is given by (\ref{proj}) with the sign in front of fermion transformation flipped.
Since the tension of these orientifolds is negative, it is bent toward internal direction and eventually should smoothly connect with each other. This can be regarded as a boundary version of the AdS soliton.

On the other hand, if we consider a system with a $O8^{-}$ and a $O8^{+}$, then a half of the original supersymmetries are preserved. In this case, since both NSNS and RR tadpole are canceled,
we do not need any D8-branes. The $Z_2$ projection for $O8^{+}$ is given by replacing
the SO projection in (\ref{proj}) with the Sp projection. In this case, we can indeed
confirm that the
remained fermion zero modes are non-chiral after we impose the two boundary conditions.
Finally if we consider two  $O8^{+}$-planes, then we need to insert 8 anti D-branes
at each boundaries. This again leads to eight chiral (complex) fermions at each of them.

Finally we would like to notice that if we take the strong coupling limit, we will find the
Horava-Witten model \cite{HW}, where the boundary $Q$ is now introduced in the
$AdS_4\times S^7/Z_k$ background of M-theory.

\section{Conclusions and Discussions}

In this paper we studied various aspects of the holographic dual of BCFT
(AdS/BCFT).
We explicitly constructed holographic duals of
conformal field theories on balls.

In AdS$_5/$BCFT$_4$, we confirmed that the coefficient of
the logarithmic term in the partition function agrees with that computed from the
conformal anomaly in the field theory. In AdS$_4/$BCFT$_3$, we noticed there is again
a logarithmic term and we found that this is associated with the boundary degrees of
freedom as in the central charge in CFT$_2$. Thus we called this a boundary central charge.
It will be an intriguing future direction to study this quantity from a field theory side.
In AdS$_3/$BCFT$_2$, the partition function is interpreted as the g-function.

Next we showed a holographic g-theorem in any dimension, which argued that the holographically
defined g-function decreases under RG flows if we assume the null energy condition.
As a special example of this, we found that the boundary central charge in BCFT$_3$
decreases under RG flows. We also pointed out this
g-theorem is closely related to the topological censorship which prohibits the existence of
AdS wormholes.

If we heat up such systems, it is natural to
 expect a Hawking-Page type phase transition. We explicitly show this transition in
 the AdS$_3/$BCFT$_2$ case. We find that as the boundary entropy gets larger, the transition
 temperature gets lower. In higher dimensions, though we found a solution in the
 low temperature phase, we did not in the high temperature phase. This leaves an
 interesting future problem.

Moreover, we presented a time-dependent example of AdS/BCFT. In this example,
the two disconnected regions at the AdS boundary are accelerated so that
they cannot communicate with each other. These two regions are connected in the
bulk AdS. We confirmed that the entanglement entropy between them is equal to the
black hole entropy after a coordinate transformation which makes the background static.

We also computed holographic one-point functions and confirmed that their scaling property is the same as what we expect from field theory calculations. Finally, we gave an example of string theory embedding of this holography, based on the type IIA AdS$_4\times$CP$^3$ geometry
with O8-planes. This is dual to a ABJM theory on an interval. The gauge anomaly cancelation
requires massless chiral fermions localized at the two dimensional boundary, being
consistent with the D-brane analysis.
In this example, even though we can change the value of
the tension $T$, it is possible that the back reactions of the D8-branes and O8-planes
break the conformal invariance. This issue deserves future study. At the same time,
it is an important future problem to find some other examples
in string theory and realize AdS/BCFT with a non-vanishing tension $T$.

\section*{Acknowledgements}
The authors are grateful to L. Giomi, M. Headrick, Y. Hikida, A. Karch, J. Maldacena, J. McGreevy, R. Myers, T. Nishioka, N. Ogawa, S. Ryu,
T. Ugajin and E. Witten for valuable comments. MF and TT are supported by World Premier International Research Center Initiative (WPI Initiative) from the Japan Ministry of Ecucation, Culture, Sports, Science and Technology (MEXT).
MF is supported by the postdoctoral fellowship program of
the Japan Society for the Promotion of Science (JSPS),
and partly by JSPS Grant-in-Aid for JSPS Fellows No.\,22$\cdot$1028.
TT thanks very much the Strings 2011 conference in Uppsala, where a part of this work
has been announced and conducted. TT and ET are
grateful to the Aspen center for physics and the Aspen workshop
``Quantum Information in Quantum Gravity and Condensed Matter
Physics,'' where a part of this work was conducted.
TT is partly supported by JSPS Grant-in-Aid for Scientific Research No.\,20740132
and by JSPS Grant-in-Aid for Creative Scientific Research No.\,19GS0219.
ET is supported by Istituto Nazionale di Fisica Nucleare (INFN) through the ``Bruno Rossi'' fellowship and by funds of the U.S. Department of Energy under the cooperative research agreement DE-FG02-05ER41360.

\appendix

\section{Calculations of Extrinsic Curvatures}\label{apext}

Here we would like to explain the calculations of extrinsic curvatures in $d+1$
dimensional spacetime. If we define $h_{\mu\nu}$ ($\mu,\nu=0,1,2,\ddd,d$)
to be the induced metric of d-dimensional
submanifold $Q$ and $n^\mu$ to be a space-like
unit normal vector on $Q$ (toward the outside direction), then we
have the relation
\be
\label{ind met}
h_{\mu\nu}=g_{\mu\nu}-n_\mu n_\nu.
\ee
The extrinsic curvature $K_{\mu\nu}$ is defined by
\be
\label{K def}
K_{\mu\nu}=h_\mu^\rho  h_\nu^\lambda \nabla_\rho n_\lambda,
\ee
where $\nabla_\rho n_\lambda$ is the standard covariant derivative in the $d+1$ dimensional
spacetime. Its trace is given by
\be
\label{K trace def}
K=g^{\mu\nu}K_{\mu\nu}=h^{\mu\nu}K_{\mu\nu}.
\ee
These expressions using the $d+1$ dimensional coordinate are degenerate because
$K_{\mu\nu}n^\nu=h_{\mu\nu}n^\nu=0$. In this convention, the boundary equation of motion (\ref{eqbein})
is equivalent to
\be
K_{\mu\nu}=(K-T)h_{\mu\nu}. \label{BOE120}
\ee
After projecting to the $d$ dimensional coordinate, we obtain $K_{ab}$ and $h_{ab}$
($a,b=0,1,2,\ddd,d-1)$.

\section{Possibility of Connected Boundary for the Annulus}
\label{appannulus}

\begin{figure}[ttt]
   \begin{center}
     \includegraphics[height=5cm]{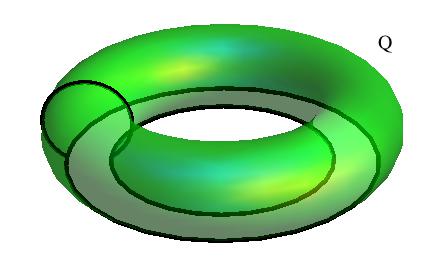}
   \end{center}
      \vspace{-.6cm}
   \caption{A connected surface in the bulk associated to the annulus.}
   \label{fig:annulusconnected}
\end{figure}

\noindent In this appendix we consider the boundary theory defined on an annulus and the connected surface $Q$ extending in the bulk shown in Fig. \ref{fig:annulusconnected} as a possible solution of (\ref{eqbein}) in Euclidean $AdS_3$ (\ref{poth}).\\
We find it convenient to parameterize this surface by two angular coordinates $(\theta,\phi)$ as follows
\begin{equation}
z\,=\,r_1 (\sin \phi+\sin \delta)
\hspace{1cm}
\tau\,=\,(r_0 + r_1 \cos \phi) \cos \theta
\hspace{1cm}
x\,=\,(r_0 + r_1 \cos \phi) \sin \theta
\end{equation}
where $0 \leqslant \theta \leqslant 2\pi$ and $-\delta  \leqslant \phi \leqslant \pi+\delta $ while $\delta \in (-\pi/2,\pi/2)$ is a parameter. The circumference in the middle of the boundary annulus has radius $r_0$ while $r_1<r_0$ is the radius of the circumference arch shown in Fig. \ref{fig:annulusconnected}. The parameter $\delta$ tells us whether the center of this circumference arch has $z>0$ ($\delta >0$) or $z<0$ ($\delta <0$).
The surface $Q$ is generated by rotating this arch around a vertical axis defined by $\tau=0$ and $x=0$. The thickness of the boundary annulus is $2 r_1 \cos\delta$.\\
The surface $Q$ is equivalently described by the following constraint
\begin{equation}
\Big(\sqrt{\tau^2+x^2}-r_0\Big)^2+(z-r_1 \sin\delta)^2 \,=\,r_1^2\,.
\end{equation}
The induced metric on $Q$ reads
\begin{equation}
h_{ab}\,=\,\frac{R^2}{z^2}
\left(\begin{array}{cc}
(r_0 +r_1 \cos \phi)^2 & 0\\
0 & r_1^2
\end{array}\right)
\end{equation}
while from the expression of $K_{ab}$ for $Q$ we find that
\begin{equation}
K_{ab}- K\, h_{ab}\,=\,\frac{R}{z^2}
\left(\begin{array}{cc}
-\,(r_0 +r_1 \cos \phi)^2 \sin \delta & 0\\
0 &
\displaystyle \frac{r_0 \sin \phi-r_1 \sin \delta \cos \phi}{r_0 + r_1 \cos \phi}\,r_1^2
\end{array}\right)\,.
\end{equation}
It is clear that we cannot solve both the two equations given by $K_{ab} -K h_{ab} = -T h_{ab}$ through the parameter $\delta$ only. Indeed, the first equation provides the tension $T=\sin\delta /R$ but then the other one leads to $r_0=0$, which makes the annulus collapse to a circle and $Q$ becomes the surface shown in Fig. \ref{figsphere}.

\section{Boundaries in Rotating BTZ Black holes}\label{aprot}

In this appendix we find a surface in the rotating BTZ satisfying the boundary equation \eqref{BOE120} by using the fact that it is locally equivalent to $AdS_3$.
The metric of the rotating BTZ is given by
\be
ds^2=\dfrac{r^2}{(r^2-r_+^2)(r^2-r_-^2)}dr^2+r^2dx^2+\Big(-\dfrac{(r^2-r_+^2)(r^2-r_-^2)}{r^2}+\dfrac{r_+^2r_-^2}{r^2}\Big)dt^2+2r_+r_-dtdx, \label{ROT121}
\ee
where we used the convention of the unit AdS radius $R=1$.

This metric is obtained by a coordinate transformation of the pure $AdS_3$ metric $ds^2=(dw_+dw_-+dz^2)/z^2$ as follows:
\begin{align}
&w_+=\sqrt{\dfrac{r^2-r_+^2}{r^2-r_-^2}}\exp((x+t)(r_++r_-))\equiv T+V, \\
&w_-=\sqrt{\dfrac{r^2-r_+^2}{r^2-r_-^2}}\exp((x-t)(r_+-r_-))\equiv -T+V, \\
&z=\sqrt{\dfrac{r_+^2-r_-^2}{r^2-r_-^2}}\exp(xr_++tr_-).
\end{align}

This transformation changes the following sphere to a surface:
\be
w_+w_-+(z-r_B\sinh(\rho_*))^2-r_B^2\cosh(\rho_*)^2=0.
\ee
 Remind that $R$ is set to be 1 and this surface is a solution of the gravity dual of BCFT.  Employing the coordinate transformation above, the surface in the rotating BTZ black hole is given by
\be
 \label{Fil06}
0\,=\,F(t,x,r) \, \equiv \,
x+ \frac{1}{r_+}\left[ \,
\rho_*+t\,r_{-}  -\log\left(\frac{r_B}{2}\right)
- \log \big(\tilde{f}(r) \big)
\right]
\ee
where
\be
\tilde{f}(r)\,\equiv\,
(e^{2\rho_\ast}-1)\sqrt{\frac{r_+^2-r_-^2}{r^2-r_-^2}}
+
\sqrt{\frac{e^{4\rho_\ast}(r_+^2-r_-^2)+2e^{2\rho_\ast}(2r^2-r_+^2-r_-^2)+r_+^2-r_-^2}{r^2-r_-^2}}\,.
\ee
Remind that for the case $r_-=0$, the rotating BTZ black hole becomes the BTZ black hole. Setting $r_-=0$, indeed, the surface (\ref{Fil06}) becomes the surface of the BTZ BH found in \eqref{solx}:
\be
x+ \frac{1}{r_+}\left[ \,
\rho_*+ -\log\left(\frac{r_B}{2}\right)
- \log \left(
(e^{2\rho_\ast}-1)\frac{r_+}{r}+
\frac{\sqrt{4e^{2\rho_\ast} r^2+(e^{2\rho_\ast}-1)^2 r_+^2 }}{r}
\right)
\right]\,=\,0\,.
\ee
 Taking the derivative and using the coordinate transformation $r\to 1/z$,
 \be
\frac{dx}{dz}= \frac{T}{\sqrt{1-T^2(1-r_+^2z^2)}}.
 \ee
In addition, the surface \eqref{Fil06} for the rotating BTZ \eqref{ROT121} is also a solution of the boundary equation.

To check this, first, the surface normal reads
\be
\hat{n}_{\mu}=\partial_{\mu}F(t,x,r),
\hspace{2cm}
n_{\mu}=\dfrac{\hat{n}_{\mu}}{\sqrt{\hat{n}_{\nu}\hat{n}^{\nu}}}
\ee
while the induced metric is defined in (\ref{ind met}).
The second Christoffel symbol for the rotating BTZ black hole is obtained as
\begin{eqnarray}
& & \hspace{-1.5cm}
\Gamma^{3}_{13}={\frac {{r}^{3}}{ \left( {r}^{2}-{{\it r_+}}^{2} \right)  \left( {r}^{2
}-{{\it r_-}}^{2} \right) }},\quad \Gamma ^3_{12}={\frac {{\it r_+}\,{\it r_-}\,r}{ \left( {r}^{2}-{{\it r_+}}^{2} \right)
 \left( {r}^{2}-{{\it r_-}}^{2} \right) }},\quad \Gamma^2_{13}= -{\frac {{\it r_+}\,{\it r_-}\,r}{ \left( {r}^{2}-{{\it r_+}}^{2}
 \right)  \left( {r}^{2}-{{\it r_-}}^{2} \right) }}, \\
 \rule{0pt}{.8cm}
& & \hspace{-1.5cm}
\Gamma ^2_{12}={\frac { \left( {r}^{2}-{{\it r_-}}^{2}-{{\it r_+}}^{2} \right) r}{
 \left( {r}^{2}-{{\it r_+}}^{2} \right)  \left( {r}^{2}-{{\it r_-}}^{2}
 \right) }},\quad \Gamma^1_{33}={\frac {{r}^{4}-{r}^{2}{{\it r_-}}^{2}-{{\it r_+}}^{2}{r}^{2}+{{\it r_+}}
^{2}{{\it r_-}}^{2}}{r}},\quad   \\
 \rule{0pt}{.8cm}
& & \hspace{-1.5cm}
\Gamma ^1_{22}-{\frac {{r}^{4}-{r}^{2}{{\it r_-}}^{2}-{{\it r_+}}^{2}{r}^{2}+{{\it r_+}
}^{2}{{\it r_-}}^{2}}{r}},\quad \Gamma ^1_{11}=-{\frac {{r}^{4}-{{\it r_+}}^{2}{{\it r_-}}^{2}}{r \left( {r}^{2}-{{\it
r_+}}^{2} \right)  \left( {r}^{2}-{{\it r_-}}^{2} \right) }}\;.
\end{eqnarray}
The extrinsic curvature are obtained by using (\ref{K def}) and (\ref{K trace def}).
After a short computation, the surface \eqref{Fil06} is found to be the solution as follows:
\be
K=2T,\quad K_{\mu\nu}-\dfrac{K}{2}h_{\mu\nu}=0.
\ee

\section{Details about the One Point Function}
\label{apsf}

In this appendix we give some technical details on the derivation of the results of the section \ref{corr section}.\\
Given (\ref{Phi7}), it is useful to recall that the modified Bessel function of the second kind $K_{\nu}(z)$  is related to the modified Bessel function of the first kind  $I_{\nu}(z)$ as follows
\be
K_{\nu}(z)=\dfrac{\pi}{2}\dfrac{I_{-\nu}(z)-I_{\nu}(z)}{\sin \nu\pi}
\hspace{1.7cm}
I_{\nu}(z)=\Big(\dfrac{z}{2} \Big)^{\nu}\sum_{n=0}^{\infty}\dfrac{(z/2)^{2n}}{n!\Gamma (\nu +n+1)}
\,.
\ee
where we have also provided the expansion of $I_{\nu}(z)$.
Let us consider the boundary condition (\ref{bcsol})
\begin{equation}
\label{bc step2}
\int_{-\infty}^{+\infty}A(k) \left[\,
ik\,K_\nu(|k| z) \cos\theta-\left(\frac{d}{2z}\,K_\nu(|k| z)+\partial_z K_\nu(|k| z)\right)\sin\theta\,
\right] e^{ikz \tan\theta} dk
\,=\,
-\frac{aR}{z^{d/2+1}}\,.
\end{equation}
By using that
\begin{equation}
\partial_z K_\nu(z) \,=\,-\frac{K_{\nu-1}(z) +K_{\nu+1}(z) }{2}
\end{equation}
and (\ref{cdef}), the boundary condition (\ref{bc step2}) reads
(we change the integration variable $\tilde{k} \equiv k z$ here)
\begin{eqnarray}
\label{bc step4}
\frac{c_\theta}{2}
\int_{-\infty}^{+\infty} |\tilde{k}|^{d/2}\bigg\{
2K_\nu(|\tilde{k}|) \cos\theta\\
& & \hspace{-2.3cm}+
\left[\,
\frac{d}{\tilde{k}}\,K_{\nu}(|\tilde{k}|)
-\frac{|\tilde{k}|}{\tilde{k}}\Big(K_{\nu-1}(|\tilde{k}|)+K_{\nu+1}(|\tilde{k}|)\Big)\right]i \sin\theta
\bigg\}\, e^{i \tilde{k} \tan\theta} d\tilde{k}
=\;i aR\,.
\nonumber
\end{eqnarray}
By exploiting the parity of the integrand, we can write (\ref{bc step4}) as follows
\begin{eqnarray}
\label{bc step5}
c_\theta
\int_{0}^{+\infty} \tilde{k}^{d/2}\bigg\{
2K_\nu(\tilde{k}) \cos\theta
\Big( e^{i \tilde{k} \tan\theta} + e^{-i \tilde{k} \tan\theta} \Big)
& &\\
& & \hspace{-6cm}+\;i \sin\theta
\left[\,\frac{d}{k}\,K_{\nu}(\tilde{k})
-K_{\nu-1}(\tilde{k})-K_{\nu+1}(\tilde{k})
\,\right]
\Big( e^{i \tilde{k} \tan\theta} - e^{-i \tilde{k} \tan\theta} \Big)
\bigg\}\, d\tilde{k} \,=\,
  2i \,aR
\nonumber
\end{eqnarray}
where we can recognize $\cos(\tilde{k} \tan\theta)$ and $\sin(\tilde{k} \tan\theta)$ in the expressions between the round brackets.
The integral in the l.h.s. of (\ref{bc step5})  clearly provides a real number, thus $c_\theta$ is purely imaginary.

Let us consider first the simplest case of $\theta=0$ i.e. $x=0$.
The condition (\ref{bc step5}) then becomes
\begin{equation}
\label{bc theta0}
2c_0 \int_0^{+\infty} k^{d/2} \,K_\nu(k) \, dk
\,=\,
i \,aR\;.
\end{equation}
By using the following integral in (\ref{bc theta0})
\begin{equation}
\label{integ1}
\int_0^{\infty}k^{\mu-1} K_\nu(k)\,dk\,=\,
2^{\mu-2} \,\Gamma\Big(\frac{\mu+\nu}{2}\Big)\,\Gamma\Big(\frac{\mu-\nu}{2}\Big)
\hspace{1.6cm}
\left\{
\begin{array}{l}
\textrm{Re}(\mu)\,>\,\textrm{Re}(\nu)\\
\textrm{Re}(\mu+\nu)\,>\,0
\end{array}\right.
\end{equation}
and also (\ref{cdef}), we get \ref{Ak0}.
In the general case of $\theta \geqslant 0$, by employing the following integrals
\begin{eqnarray}
\int_0^{\infty}k^{\mu-1} K_\nu(k) \cos(bk)\,dk
&=&
2^{\mu-2} \,\Gamma\Big(\frac{\mu+\nu}{2}\Big)\,\Gamma\Big(\frac{\mu-\nu}{2}\Big)
\, F\Big(\frac{\mu+\nu}{2} , \frac{\mu-\nu}{2}; \frac{1}{2} ; \,-\,b^2 \Big)
\hspace{1cm}\\
\rule{0pt}{.6cm}
\int_0^{\infty}k^{\mu-1} K_\nu(k) \sin(bk)\,dk
&=& \\
& &\hspace{-2cm}=\;
2^{\mu-1} b\, \,\Gamma\Big(\frac{\mu+\nu+1}{2}\Big)\,\Gamma\Big(\frac{\mu-\nu+1}{2}\Big)
\, F\Big(\frac{\mu+\nu+1}{2} , \frac{\mu-\nu+1}{2}; \frac{3}{2} ; \,-\,b^2 \Big)
\nonumber
\end{eqnarray}
 the boundary condition (\ref{bc step5}) becomes
 \begin{eqnarray}
 \label{bc step7}
\frac{2^{d/2}}{\cos\theta} \, \Gamma(\alpha_+)\,\Gamma(\alpha_-)
\,\bigg\{
\cos^2\theta\;
F\big(\alpha_+ , \alpha_-; 1/2 \,; -\,v^2 \big) \\
& & \hspace{-8cm}
-\,\sin^2\theta
\Big[(2\alpha_- -1)
\, F\big(\alpha_+ , \alpha_-; 3/2 \,; - \,v^2 \big)
-2 \alpha_-
\, F\big(\alpha_+ , \alpha_- +1; 3/2 \,; - \,v^2 \big)
\Big] \bigg\}
\,=\,\frac{i\,aR}{c_\theta}
\nonumber
\end{eqnarray}
where
\begin{equation}
\alpha_+ \,\equiv \, \frac{d/2+\nu+1}{2}
\hspace{1.5cm}
\alpha_- \,\equiv \, \frac{d/2-\nu+1}{2}
\hspace{1.5cm}
v\,\equiv\,\tan\theta\;.
\end{equation}
By employing the following identity of the hypergeometric function
\begin{equation}
(c-1)\,  F(a,b;c-1;z) -b\, F(a,b+1;c;z)
+(b-c+1)\,  F(a,b;c;z)\,=\,0
\end{equation}
the boundary condition (\ref{bc step7}) simplifies to
\begin{equation}
2^{d/2}\, \Gamma(\alpha_+)\,\Gamma(\alpha_-)
\sqrt{1+v^2}\, F\big(\alpha_+ , \alpha_-; 1/2 \,; -\,v^2 \big)
\,=\,\frac{i\,aR}{c_\theta}\;.
\end{equation}
From this equation we obtain $c_\theta$, given in (\ref{ctheta}) . For $\theta=0$ we recover (\ref{Ak0}) as special case.

\section{Discrete Symmetries of ABJM Theory}\label{discs}

Here we summarize the discrete symmetries of the ABJM theory. The two $U(N)$ gauge
fields are denoted by $A^{(1)}_\mu$ and $A^{(2)}_\mu$. The matter fields consists
of bi-fundamental four complex scalars $X^I$ and bi-fundamental four Dirac fermions $\Psi^I$ ($I=1,2,3,4$). See also \cite{ABJM,Rey} for earlier discussions on the discrete symmetries.
The charge conjugation symmetry $C$ is given by
\ba
&& A^{(1,2)}_{\mu}(t,x,y)\to -A^{(1,2)}_{\mu}(t,x,y)^*,\no
&& X^I(t,x,y)\to X^I(t,x,y)^{*},\no
&& \Psi^I(t,x,y)\to C\cdot\Psi^I(t,x,y)^{*},
\ea
where $T$ denotes the transposition of the matrix and $C$ denotes the matrix which describes the parity transformation of three dimensional fermions. Notice that the two gauge groups are not
interchanged in this $C$ transformation. The parity symmetry $P$ is given by
\ba
&& A^{(1,2)}_{t,x}(t,x,y)\to -\left(A^{(2,1)}_{t,x}(t,x,-y)\right)^T,\no
&& A^{(1,2)}_{y}(t,x,y)\to \left(A^{(2,1)}_{y}(t,x,-y)\right)^T,\no
&& X^I(t,x,y)\to X^I(t,x,-y)^{T},\no
&& \Psi^I(t,x,y)\to P\cdot\Psi^I(t,x,-y)^{T},
\ea
where the two gauge groups are exchanged. Finally, the time reversal
symmetry $T$ is found to be
\ba
&& A^{(1,2)}_{t}(t,x,y)\to \left(A^{(2,1)}_{t}(-t,x,y)\right)^T,\no
&& A^{(1,2)}_{x,y}(t,x,y)\to -\left(A^{(2,1)}_{x,y}(-t,x,y)\right)^T,\no
&& X^I(t,x,y)\to X^I(t,x,-y)^{T},\no
&& \Psi^I(t,x,y)\to T\cdot\Psi^I(t,x,-y)^{T},
\ea
where the two gauge groups are again exchanged. Remember that a standard Chern-Simons gauge theory
with a single gauge group does not have either parity or time reversal symmetry~\cite{Dunne:1998qy}.

For a Dirac fermion description with the signature $\eta_{\mu\nu}=(1,-1,-1)$, the gamma matrices can be defined by $\gamma_0=\sigma_1, \gamma_1=-i\sigma_2, \gamma_2=i\sigma_3$ ($\sigma_{1,2,3}$ are the Pauli matrices). In this convention, the matrix $C,P,T$ is given by
$C=i\gamma_1, P=i\gamma_2$ and $T=i\gamma_1$.

\end{document}